\documentclass[aps,prl,twocolumn,showpacs,superscriptaddress,floatfix]{revtex4-1}
\usepackage[utf8]{inputenc}
\usepackage{ucs}
\usepackage{natbib}
\usepackage{graphicx}  % needed for figures
\usepackage{bm}        % for math
\usepackage{amssymb}   % for math
\usepackage{amsmath}
\usepackage{color}
\usepackage{CJK}
\usepackage{times}
\usepackage{verbatim}
\usepackage{mathrsfs}
\usepackage{hyperref}
\usepackage{ulem}

\hypersetup{colorlinks = True, urlcolor=blue, linkcolor=blue, citecolor=blue}
\newcommand{\cred}{\color{black}}

\usepackage{cancel}
\usepackage{comment}

\def\nio{Nd$_2$Ir$_2$O$_7$}
\def\rio{R$_2$Ir$_2$O$_7$}
\def\eio{Eu$_2$Ir$_2$O$_7$}

\begin{document}

%\title{Time reversal and lattice symmetry breaking in Nd$_2$Ir$_2$O$_7$ observed by Raman scattering}

\title{Ramification  of complex magnetism  in Nd$_2$Ir$_2$O$_7$ observed by Raman scattering spectroscopy}
\author{Yuanyuan Xu}
\affiliation{Institute for Quantum Matter and Department of Physics and Astronomy, Johns Hopkins University, Baltimore, Maryland 21218, USA}

\author{Yang Yang}
\affiliation{School of Physics and Astronomy, University of Minnessota, Minneapolis, MN 55455, USA}

\author{J\'er\'emie Teyssier}
\affiliation{Department of Quantum Matter Physics, University of Geneva, Geneva, Switzerland}

\author{Takumi Ohtsuki}
\affiliation{Institute for Solid State Physics, University of Tokyo, Kashiwa, Chiba 277-8581, Japan}

\author{Yang Qiu}
\affiliation{Institute for Solid State Physics, University of Tokyo, Kashiwa, Chiba 277-8581, Japan}

\author{Satoru Nakatsuji}
\affiliation{Institute for Quantum Matter and Department of Physics and Astronomy, Johns Hopkins University, Baltimore, Maryland 21218, USA}
\affiliation{Institute for Solid State Physics, University of Tokyo, Kashiwa, Chiba 277-8581, Japan}
\affiliation{Department of Physics, University of Tokyo, Bunkyo-ku, Tokyo 113-0033, Japan}
\affiliation{Trans-scale Quantum Science Institute, University of Tokyo, Bunkyo-ku, Tokyo 113-0033, Japan}
\affiliation{CREST, Japan Science and Technology Agency, Kawaguchi, Saitama 332-0012, Japan}
\author{Dirk van der Marel}
\affiliation{Department of Quantum Matter Physics, University of Geneva, Geneva, Switzerland}

\author{Natalia  B. Perkins}
\affiliation{School of Physics and Astronomy, University of Minnessota, Minneapolis, MN 55455, USA}

\author{Natalia Drichko}
\affiliation{Institute for Quantum Matter and Department of Physics and Astronomy, Johns Hopkins University, Baltimore, Maryland 21218, USA}
\email{Corresponding author. Email:drichko@jhu.edu}

\date{\today}

\maketitle

\textbf{
Using Raman scattering spectroscopy, we uncover a complex magnetic behavior of  \nio, which stands  out  among magnetic pyrochlores  by   the lowest temperature of the all-in-all-out (AIAO) Ir moments ordering ($T^\mathrm{N}_{\small\rm{Ir}}=33$~K) and the highest  temperature at which  AIAO order of  rare-earth Nd ions is detected  ($T^\mathrm{*}_{\small\rm{Nd}}$=15~K). 
{\cred Our findings suggest that in the temperature range between 15~K and 33~K, Nd magnetic moments exhibit strong fluctuations, possibly originating from spin ice behavior.}
 This complex behavior emerges from the interplay of strong spin-orbit coupling, electronic correlations, and geometric frustration
on two magnetic pyrochlore sublattices of Nd  and Ir ions.  The ordering of  Ir magnetic moments  is accompanied by an appearance of  one-magnon  Raman modes at 26.3 and 29.6 meV  compatible with the AIAO  order and  of a  broad mode at 14 meV, {\cred which could be associated  with spinon continuum arising from Nd spin ice fluctuations}. While two one-magnon excitations show minimal temperature evolution with decreasing temperature,
the 14 meV mode shifts to higher frequencies as the temperature approaches a crossover to Nd AIAO  order, broadens,   and  disappears below 15~K. An additional two-magnon excitation of  the AIAO  Nd order at around 33 meV appears in the spectra at low temperatures. These rather high energies of  magnetic excitations  of Nd  moments make \nio\  a particularly attractive playground to study the rare-earth magnetism on the pyrochlore lattice.
}

\bigskip

\section{Introduction}

Topological magnets provide a fertile platform to study novel phenomena through their nontrivial topological magnetic excitations
~\cite{Nagaosa2013,Fert2013,Witczak2014,Takagi2019,Broholm2020,nakatsuji_annurev-conmatphys-031620-103859}. Among them, magnetic Weyl semimetals
 with time reversal symmetry  (TRS) breaking
attract much attention due to their striking properties
and a potential for various  applications ~\cite{Armitage2018,Nakatsuji2015,Sakai2018,Liu2018,Liu2019a,Belopolski2019}.
Pyrochlore iridates of general formula \rio~ with R being a
rare earth element  Y, Eu, Nd, Sm, or Pr~\cite{Nakatsuji2006,Machida2010,Pesin2010,Witczak2014,Cheng2017,Ohtsuki2019} were among the first materials predicted to host Weyl fermions ~\cite{Wan2011}.
In these materials,  a Weyl semimetal state can be brought about by a splitting of the quadratic band touching node into pairs of Weyl nodes either under TRS breaking produced  by magnetic ordering of Ir magnetic moments~\cite{Wan2011,Witczak2012}   or by a loss of inversion symmetry center \cite{Bzdusek2015}. Particularly, AIAO ordering of Ir$^{4+}$ magnetic moments  below $T_\mathrm{N}^{\small\rm{Ir}}=$ 33~K~\footnote{Publications on \nio\ show some spread in the $T_\mathrm{N}^{\small\rm{Ir}}$ values from 32~K to as high as 37~K~\cite{Wang2020}. The reason for the discrepancy between the temperature obtained from different crystals and different measurements could be extreme sensitivity of \nio\ properties to Ir/Nd stoichiometry.   According to Ref.~\cite{Nakayama2016} the transition in resistivity shifts down  to 25 K with 1\% off-stoichiometry ratio of Ir/Nd.}, which is the lowest temperature of the  magnetic ordering  of Ir$^{4+}$ moments
among pyrochlore iridates~\cite{Tomiyasu2012,Guo2013,Asih2017,Wang2020},
preserves cubic symmetry but breaks TRS \cite{Matsuhira2011}.
 There is a set of circumstantial evidence, demonstrating both quadratic band touching in Nd$_2$Ir$_2$O$_7$ at the $\Gamma$-point
in the paramagnetic high-temperature regime \cite{Ueda2012} and signatures of a magnetic Weyl semimetal associated with TRS breaking due to magnetic ordering of Ir moments below $T_\mathrm{N}^{\small\rm{Ir}}$  ~\cite{Kondo2015,Wang2020,Ueda2018}.
Recently we demonstrated that electronic Raman scattering reveals the quadratic bands above $T^\mathrm{N}_{\small\rm{Ir}}$ = 33~K and the linear dispersion of Weyl nodes below  $T^\mathrm{N}_{\small\rm{Ir}}$
~\cite{Nikolic2022}.

In \nio\  not only Ir$^{4+}$ but also Nd$^{3+}$  ions are magnetic with $J = 9/2$.   Their moments gradually order into AIAO state, 
which was experimentally observed below the crossover temperature  $T^\mathrm{*}_{\small\rm{Nd}}$=15~K, as reported in previous studies  \cite{Guo2013,Guo2016,Tomiyasu2012}. This temperature
stands out as an order of magnitude higher than the ordering temperatures for other Nd-containing pyrochlores \cite{Xu2015,Mauws2021,Ku2018,Dirk2022}, suggesting  a strong coupling  between   Nd$^{3+}$   and Ir$^{4+}$ magnetic moments. 
Specifically, the AIAO order of Ir moments
 provides  local magnetic field~\cite{Tian2016,Chen2012},
 gradually inducing the alignment of Nd moments into the AIAO order as the temperature decreases.
 Similar observations have been  also reported in several pyrochlore iridates, {\cred such as Ho$_2$Ir$_2$O$_7$, Tb$_2$Ir$_2$O$_7$  and Dy$_2$Ir$_2$O$_7$} \cite{Lefran2015,Lefrancois2017,Guo2017,Cathelin2020}, where both the ordering of Ir moments and the subsequent ordering of rare-earth moments have been observed at relatively high temperatures, again driven by the local field of Ir.

{\cred In this work, we suggest that the energy scale of magnetic excitations within the Nd subsystem may point to a notable renormalization of the magnetic interactions between Nd$^{3+}$ ions. This renormalization could stem from either additional superexchange pathways involving partially filled Ir${}^{4+}$ ions or  from
low-energy electronic excitations of the low-temperature semimetallic state \cite{Tian2016, Nikolic2022}, or both,
 and  tends to  the emergence of spin-ice behavior in the Nd subsystem in intermediate temperatures.
}
The complex interplay between renormalized exchange couplings of Nd moments, the local field 
exerted on Nd moments from the ordered AIAO state of Ir moments, and various factors such as strong spin-orbit coupling (SOC), electronic and magnetic correlations, and band topology collectively determine the unique low-temperature magnetic properties of \nio.
\begin{figure*}
	\centering
	\includegraphics[width=\linewidth]{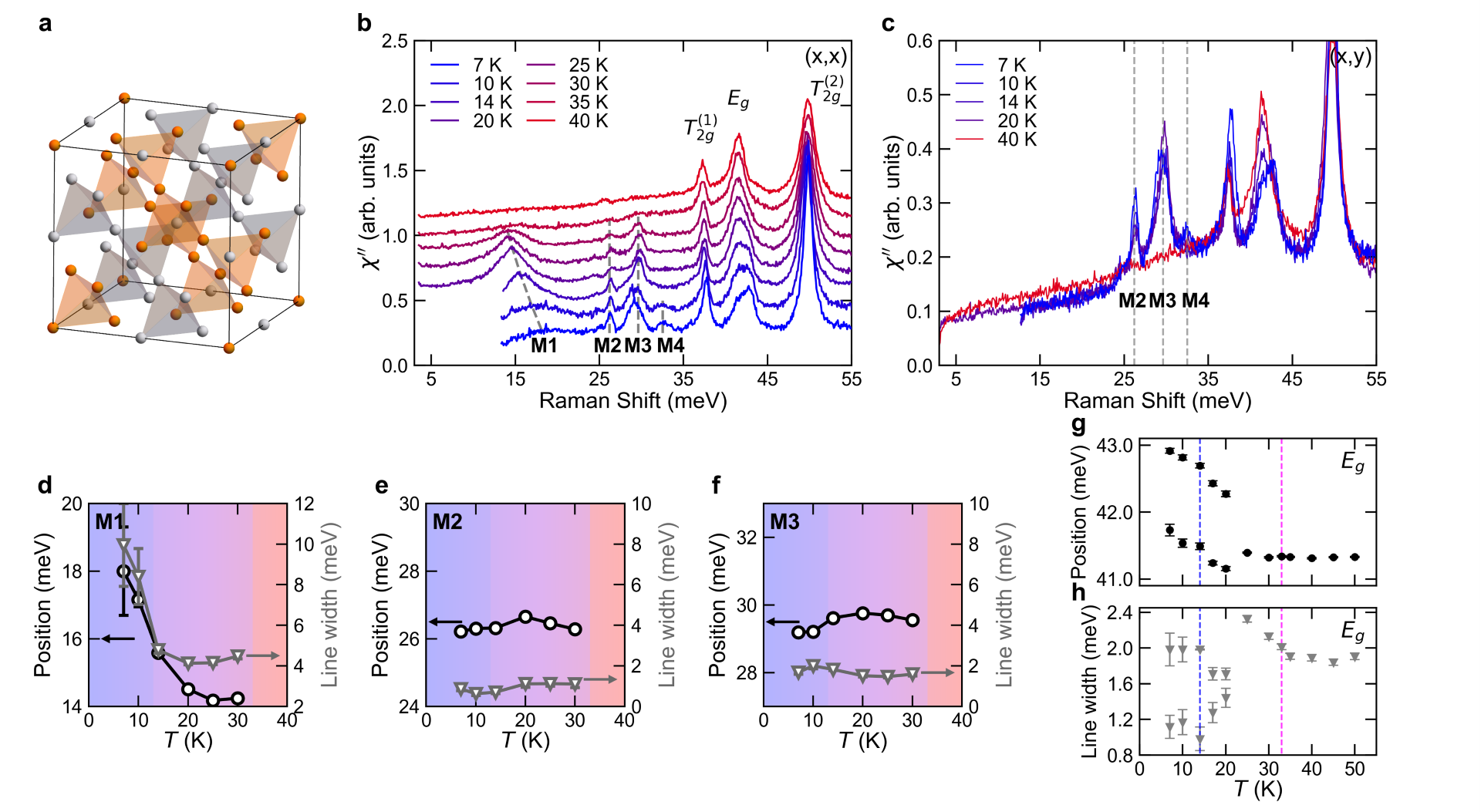}
	\caption{ {\bf a} Schematic  structure of  the pyrochlore \nio, with Nd ions shown in orange and Ir ions shown in grey.
 {\bf b-c} Temperature dependence of Raman scattering spectra of \nio\ in the spectra range between 3 and 40 meV, at temperatures between 40~K to 7~K. Spectral range below 20~K is limited by 12 meV: {\bf b} $(x,x)$ polarization, $A_{1g} + E_g + T_{2g}$ scattering channels. Note a shift of $\chi''(\omega)$ along $Y$ axis for clarity; {\bf c} $(x,y)$ polarization $E_g +T_{2g}$ scattering channels.  Magnetic excitations discussed in the text are labeled M1-M4; {\bf d-f.} Temperature dependence of the position and line width of magnetic excitations M1, M2, and M3 peaks measured in the $(x,x)$ polarization.
  The background colors, pink, magenta, and blue,
  mark,  correspondingly, the temperature regions of  the  paramagnetic phase, Ir AIAO ordered phase, Ir and Nd AIAO ordered phase.{\bf g-h} Temperature dependence of positions   and line width  of the $E_g$ phonon and its' two components observed at lower temperatures.}
	\label{fig:fig1}
\end{figure*}
We present Raman scattering data  and  model calculations,  which together identify a set of magnetic excitations  in \nio\ originated from strongly interacting Ir and Nd magnetic subsystems.  We identify  the one-magnon excitations  from the AIAO ordering of Ir moments appearing
below $T^\mathrm{N}_{\small\rm{Ir}}=33$~K, and a band of two-magnon excitations of the AIAO order of Nd moments
below  the crossover temperature $T^\mathrm{*}_{\small\rm{Nd}}$=15~K. 
In the  temperature range just below $T^\mathrm{N}_{\small\rm{Ir}}=33$~K
we observe a broad Raman excitation at about 14 meV, which shows unconventional temperature behavior, that allows us to associate it with the collective fluctuations of Nd moments, which   {\cred might be potentially understood as 
as spinon excitations  of  a quantum spin ice (QSI) phase ~\cite{Ross2011,Savary2012,Gingras2014,Hao2014,Fu2017}. }

\section{Results}

Raman scattering spectra of \nio\ in the temperature range from 50~K down to 7~K and spectral range between 3 and 40~meV were  measured in $(x,x)$  ($A_{1g} + E_g + T_{2g}$ scattering channels) and $(x,y)$ ($E_g + T_{2g}$ channels) polarizations in the [111] crystal plane (see  Fig.~\ref{fig:fig1} ({\bf b,c})).
The spectroscopic response of \nio\  and its temperature  evolution
are very rich, reflecting the complexity of the material. The spectra of \nio\ in the paramagnetic semimetallic phase above 33~K show narrow intense features of phonons  above 35 meV  superimposed on a broad electronic scattering continuum. The  observed phonon spectrum is in agreement with previously observed results from Nd$_2$Ir$_2$O$_7$ and other pyrochlore iridates~\cite{Hasegawa2010,Ueda2019,Nguyen2021,Xu2022phonons}. Detailed discussions and assignments of the  phonons can be found in the SI.  The  crystal electric field (CEF) excitations reported in previous studies of
   \nio\ \cite{Watahiki2011,Xu2015,Xu2021cef,Xu2022phonons}
are also expected to manifest in the low-temperature Raman spectrum.  Indeed, at 35 K we observed a very weak excitation at 25 meV (see Fig.~\ref{fig:fig1} b,c and Fig.~\ref{fig:spinon}g), see  Section S1.3 of SI for more information.  However, the overlap of  
this CEF excitation with the   one magnon excitations M2 and M3,  appearing in the spectra below 33 ~K, poses a challenge in studying its temperature evolution.

A phase transition at $T^\mathrm{N}_{\small\rm{Ir}}$ = 33~K is manifested by an appearance of a number of new excitations in the spectra. Excitations at 26.3 and 29.6 meV (M2 and M3) appear in both polarization channels at temperatures slightly above $T^\mathrm{N}_{\rm{Ir}}$, increase in intensity below the transition, and show small changes in the position and width on cooling. This temperature  evolution of  M2 and M3 is consistent with the expected behavior of spin wave modes above a well-defined long-range magnetic order. Note that two one-magnon  excitations at similar frequencies, 28.6 meV and 34.4 meV, were also  observed Y$_2$Ir$_2$O$_7$ \cite{Nguyen2021}.

A pronounced broad mode (M1) emerges below 33~K at 14~meV. This mode is only seen in  the $(x,x)$ channel, as depicted in Fig.\ref{fig:fig1}{b}. This mode  gains intensity  on cooling the sample down to approximately 20~K. 
As the temperature continues to decrease, the mode undergoes a continuous  shift towards higher energies, and its  width increases, as illustrated in Fig.\ref{fig:fig1}{d}. 
However, as the temperature falls below  $T^\mathrm{*}_{\small\rm{Nd}}$ = 15 K, this robust feature  gradually  disappears: while at 7K, the continuum, although of
significantly diminished in intensity, remains discernible
but at 5 K, as can be seen from the non-polarized Raman spectra
 presented in Fig.\ref{fig:spinon}{g}, it totally disappears. 
Lastly, below 15 K, a weak  peak feature (M4)  emerges in both scattering channels, with an energy of approximately 33 meV.

The absence of this strong feature at 14~meV (M1) in the $(x,y)$ scattering channel allows us to follow the change of the continuum of electronic scattering through the phase transition at $T^\mathrm{N}_{\small\rm{Ir}}$ = 33~K. The interpretation of the frequency dependence of the electronic continuum in terms of the interband excitations between quadratic bands ($T > T^\mathrm{N}_{\small\rm{Ir}}$) and Weyl bands ($T < T^\mathrm{N}_{\small\rm{Ir}}$) is discussed elsewhere~\cite{Nikolic2022}.
 
The temperature dependence of the phonons typically allows to uncover the lattice response to the magnetic ordering and changes in the electronic structure. The largest changes are observed for $E_g$ phonon at 42~meV (Fig~\ref{fig:fig1} {g,h}).  The changes onset at $T^\mathrm{N}_{\small\rm{Ir}}$ = 33~K marked by the red dashed line in the figure, when the splitting of the E$_g$ phonon is too small to resolve (Fig~\ref{fig:fig1} {g). Below 20K, where the splitting of 2 meV between the components is larger than the line width, we can clearly resolve the doublet. The spectral weight is shifted to the higher frequency component on further cooling [Fig.~\ref{fig:fig1}({b,g})] \footnote{$E_g$ mode is fit by single- and double-Lorentzian functions for the spectra above and below 20~K.}.}

 \section{Discussion}

To understand the   Raman spectra of \nio\ below 33 K, we start with analyzing the information on its  magnetic response.
The data on the  low-temperature magnetic order of the Nd and Ir sublattices were obtained by the  neutron diffraction \cite{Tomiyasu2012,Guo2016} and muon-spin relaxation ($\mu$SR) experiments \cite{Disseler2012,Guo2013,Asih2017}.  
While there are discrepancies in  the exact size of magnetic moment of Nd$^{3+}$,  
all these studies agree on
 the AIAO at 2~K. Moreover, there is a consensus that the order starts developing already below 15~K, suggesting 
the appearance of a local
magnetic field at the Nd$^{3+}$  moments, probably caused by the
magnetic ordering of Ir.
The magnetic structure of the Ir$^{4+}$ sublattice is hard to probe by neutrons,
however the combined evidence from the resonant X-ray scattering and $\mu$SR  spectroscopies  of  pyrochlore iridates including \nio\  suggests that Ir$^{4+}$ moments are ordered in the AIAO, parallel to  the surrounding net  of Nd$^{3+}$ moments~\cite{Sagayama2013, Disseler2014,Guo2013,Asih2017}.

\begin{figure}
	\centering
	\includegraphics[width=\linewidth]{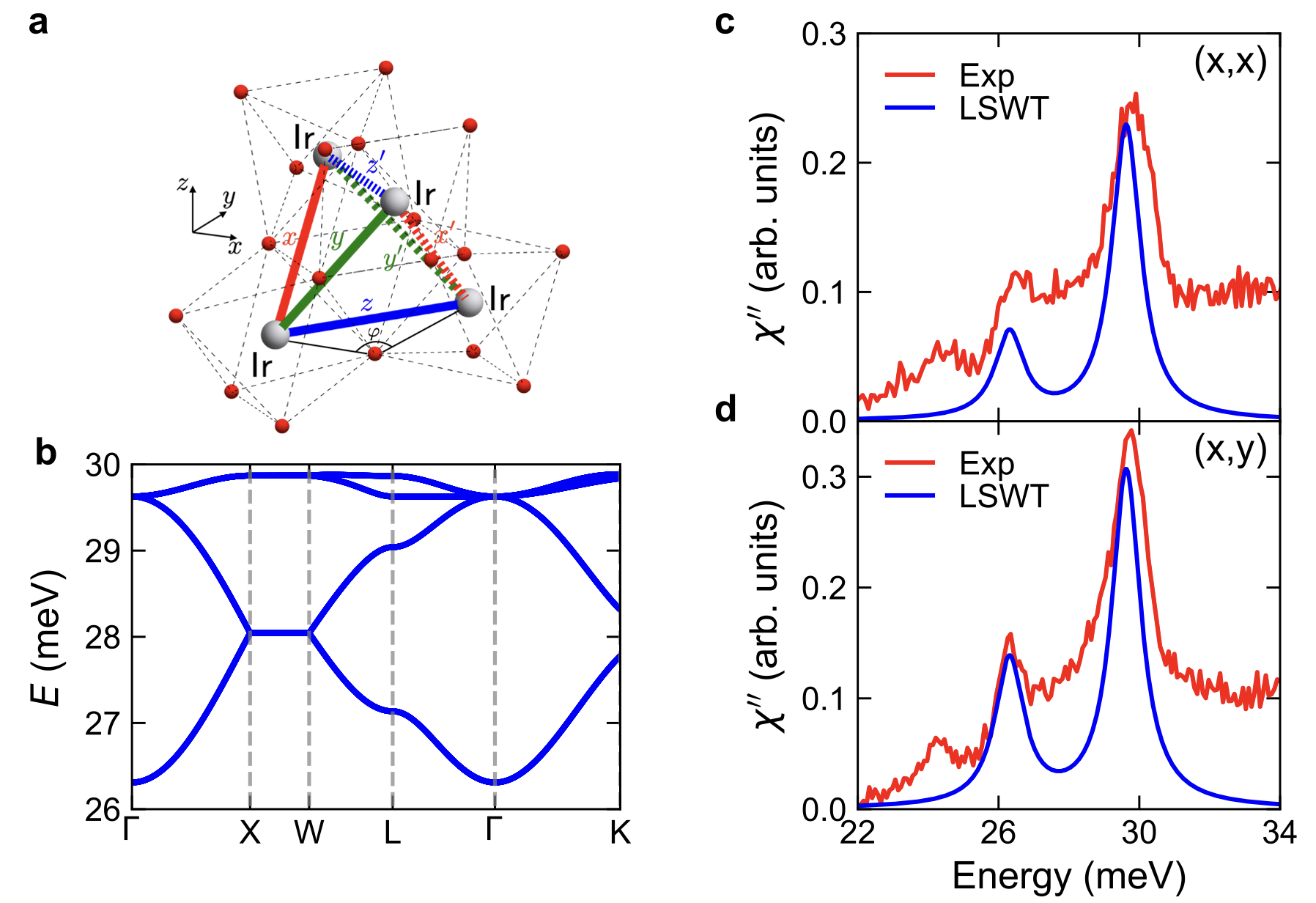}
	\caption{{\bf a.} A tetrahedron formed by $\mathrm{Ir}^{4+}$ ions: six different types of bonds  are denoted by $x$, $y$, $z$, $x'$, $y'$, and $z'$. The surrounding oxygen ions of two neighboring $\mathrm{Ir}^{4+}$ ions form two tilted corner-sharing octahedra. The angle of Ir-O-Ir is denoted by $\varphi$. {\bf b.} Magnon band structure for the AIAO  order of Ir moments obtained by the linear spin wave calculations.   {\bf c-d.} Comparison of the experimental data ($T = 20$~K, red line)  with the computed  one-magnon  Raman  response (blue line).    {\bf c.} Calculation is done for the parallel channel with arbitrary incoming light polarization in the $[111]$ plane  with
 $\mathbf{e}_\mathrm{in}^{||}=\mathbf{e}_\mathrm{out}^{||}=(0.80,-0.26,-0.54)$ for the parallel channel. {\bf d.}  Calculation is done for the cross polarized channel with  $\mathbf{e}_\mathrm{in}^\perp=\mathbf{e}_\mathrm{in}^{||}$, $\mathbf{e}_\mathrm{out}^\perp=(-0.16,0.77,-0.61)$.}
	\label{fig:magnon}
\end{figure}

{\it Magnetic excitations of Ir$^{4+}$ ions.}--The  M2 and M3 modes appear at $T^\mathrm{N}_{\small\rm{Ir}}$ and  are likely to originate from the one-magnon excitations of AIAO ordered Ir$^{4+}$ magnetic moments.
 To obtain the  magnon spectrum,  we first derive the superexchange interactions between
 effective spin-$1/2$ pseudospins of Ir$^{4+}$ ions assuming the perfect octahedral oxygen environment on the pyrochlore lattice and following the steps outlined in \cite{Sizyuk2014,Perkins2014}. The resulting superexchange Hamiltonian takes the following form:
\begin{align}
    \mathcal{H}_{\small{\mathrm{Ir}-\mathrm{Ir}}}
    =&\sum_{\langle ij\rangle_\nu}\Bigl[ J\,\mathbf{S}_i\cdot\mathbf{S}_j+K\, S_i^{\alpha_\nu}S_j^{\alpha_\nu}\nonumber\\&+\sigma_\nu\Gamma_{ij}\,(S_i^{\beta_\nu} S_j^{\gamma_\nu}+ S_i^{\gamma_\nu} S_j^{\beta_\nu})+\mathbf{D}_{ij}\cdot(\mathbf{S}_i\times\mathbf{S}_j)\Bigr],
    \label{eqn:Ir-Ir}
\end{align}
where $\nu$ determines the type of the bond [Fig.~\ref{fig:magnon}{\bf a}], and we have
$(\alpha_\nu,\beta_\nu,\gamma_\nu)=(x,y,z),\, (y,z,x),\, (z,x,y)$ for $\nu\in (x,x'),\, (y,y'),\,(z,z'),$ respectively. The prefactor $\sigma_\nu$ equals +1  for $\nu\in (x,y,z)$ and -1 for $\nu\in (x',y',z')$. The form of (\ref{eqn:Ir-Ir}) is also in agreement with the superexchange Hamiltonian obtained from the symmetry consideration in \cite{Witczak2012}.
 Given the bond-dependent anisotropy of the Hamiltonian and the non-collinear nature  of the AIAO order,  the one-magnon  response  is expected to dominate the low-energy Raman spectrum ~\cite{Yang2021}.  The AIAO state on the pyrochlore lattice gives rise to two magnon modes at center of the  Brillouin zone (BZ) $\mathbf{k}=0$: one is non-degenerate and the other is three-fold degenerate with the degeneracy protected by the symmetry of the AIAO state on the pyrochlore lattice~\cite{Hwang2020}.
The calculated linear spin wave spectra provide a good agreement with the experimental data  with the following set of parameters: $(J,K,\Gamma,D)=(6.1,\,-5.4,\,3.0,\,4.1)\,\text{ meV}$, producing  $\Gamma$-point one magnon modes  at  26.3 and 29.6 meV (are shown in Fig.~\ref{fig:magnon}{b}).
  We computed the one-magnon Raman response within the Loudon-Fleury approach~\cite{Fleury1968}, in which
  the Raman operator mirrors the processes governing exchange interactions but with virtual electron hopping being assisted by photons. It can be written as
$
    \mathcal{R}_{\small{\mathrm{Ir}}}=\sum_{\langle ij \rangle}(\mathbf{e}_\mathrm{in}\cdot\mathbf{r}_{ij})(\mathbf{e}_\mathrm{out}\cdot\mathbf{r}_{ij})\,\mathcal{H}_{\small{\mathrm{Ir}},ij}
$,
where $\mathbf{e}_\mathrm{in}$($\mathbf{e}_\mathrm{out}$) is incoming (outgoing) polarization of  light and $\mathbf{r}_{ij}$ denotes the vector connecting site $i$ and $j$ of Ir ions. 
Traditionally, it was believed that the Loudon-Fleury
response results  primarily in the  two-magnon scattering with $\Delta S^z=0$, involving the creation or destruction of a pair of magnons. However, this perspective is not universally applicable. It was recently shown \cite{Yang2021}
that  in strongly correlated Mott insulators with spin-orbit coupling and bond-dependent anisotropic interactions, a one-magnon response naturally emerges. This is also true here.

Our results for the parallel and the cross channels reproduce well the two bands M2 and M3 (Fig.~\ref{fig:magnon}{c-d}).
As a consequence of the 3:1 ratio of degeneracies of the one-magnon excitations at the $\Gamma$-point, the
computed intensity of the 29.6 meV one-magnon peak is higher than of the 26.3 meV peak, in agreement with the experiment. Moreover, the one-magnon peak M2 corresponds to the 1D irreducible corepresentation (for simplicity, we still refer them as ``irrep") of $m\bar{3}m'$ magnetic point group, which is derived from the 2D irrep $E_g$ of the paramagnetic group.

A continuum of two-magnon excitations expected within the Loudon-Fleury theory is absent in the Raman spectra of \nio\  (it was also not seen  in \eio~\cite{Ueda2019}) despite the flat magnon bands ( Fig.~\ref{fig:magnon}{b}) which would result in a peak at around 65 meV. This energy range  overlaps with the interband excitations \cite{Nikolic2022}, which can mask the observation of the two-magnon band.

 {\it Magnon-spinon dichotomy of Nd$^{3+}$ excitations.}--
 The M1 excitation at 14 meV appears below $T^\mathrm{N}_{{\mathrm{Ir}}}$  and remains relatively sharp only in a narrow temperature region (Fig.~\ref{fig:fig1}{d}). We can exclude such  origin of M1  as phonons or crystal field excitations of Nd$^{3+}$. The lowest crystal field excitation at around 25 meV is observed in the Raman spectra above T$_N^{Nd}$  (see Fig. 3 g) and overlaps with M2 magnon at lower temperatures.
 By selection rules, it is expected to appear in both (x,x) and (x,y) scattering channels (see detail discussion in SI).   Furthermore, the significant width and pronounced temperature dependence of M1 cannot be explained by either crystal field excitations or phonons~\cite{Xu2022phonons}.
Thus  we can assume that the M1 mode at 14 meV  is of magnetic origin.
However, it cannot be attributed to the magnetic moments of the Ir ions in the system.   The one-magnon excitation  M2 and M3  of Ir AIAO order exhibit only minimal modification below 33 K.  M1 excitation shows temperature and polarization dependence very  distinct from these one-magnon Ir modes.
Moreover, M1 mode gradually disappears  below the crossover temperature into  the AIAO order of Nd$^{3+}$ magnetic moments. This points out that  M1 originates from the  excitations of the disordered phase of Nd$^{3+}$ moments, while the large width suggesting a continuum of excitations.

\begin{figure*}
	\centering
	\includegraphics[width=\linewidth]{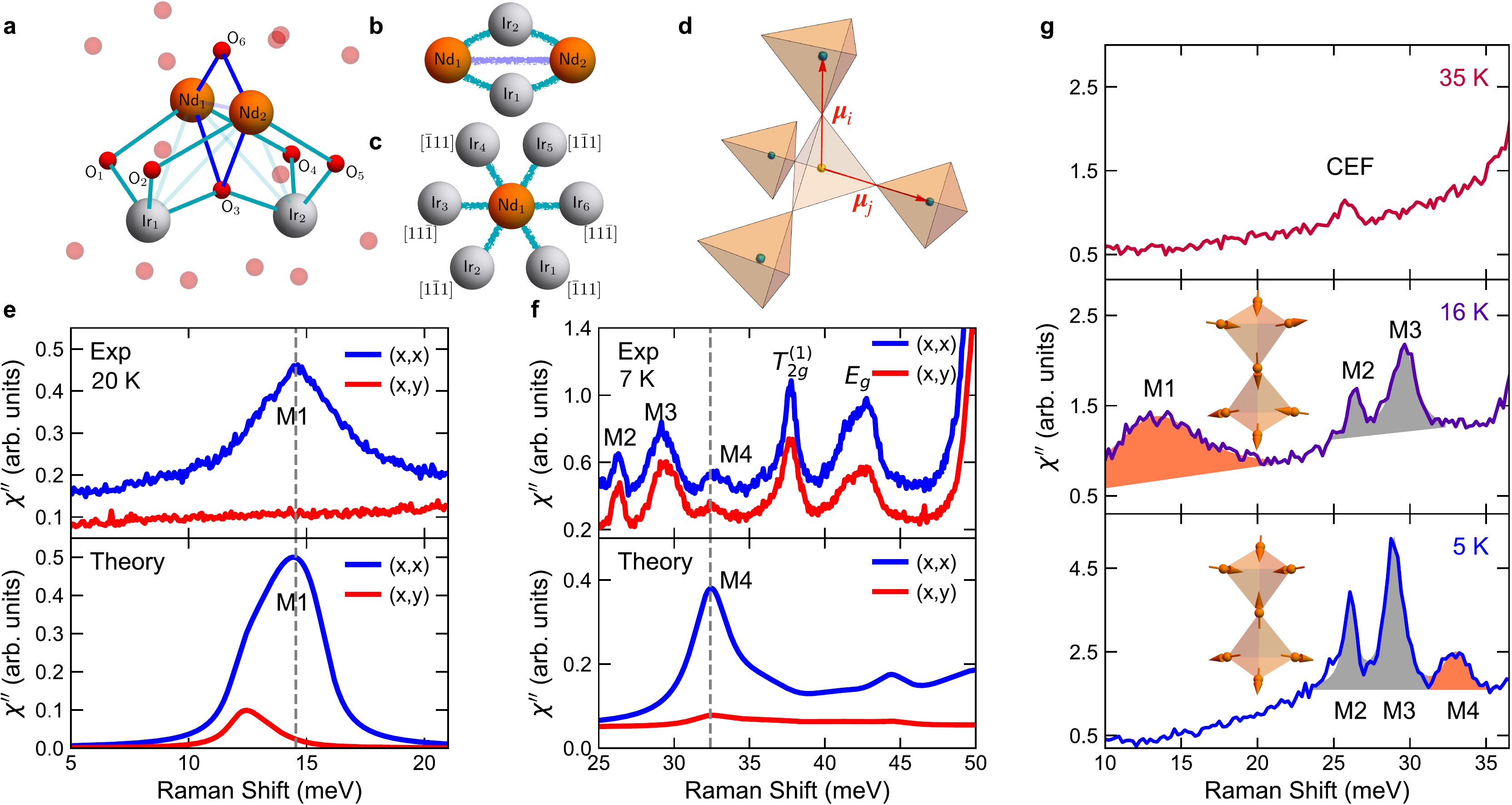}
	\caption{{\bf a.} Local environment of Nd-Nd superexchange interaction, blue bonds denote the original superexchange paths without Ir ions and teal bonds indicate additional superexchange paths involve Ir ions. {\bf b.} Schematic representation of Nd-Nd superexchange interaction, the Nd-Nd superexchange (blue) is renormalized by the additional superexchange paths involving Ir ions (teal). {\bf c.} Schematic representation of a Nd ion surrounded by its six nearest neighboring Ir ions, where the directions of the AIAO-ordered Ir moments are given in the global reference frame. 
    {\bf d.} Spinons are created on the diamond lattice (A and B sublattices of Nd tetrahedra are shown in light and dark orange, respectively) formed by the centers of Nd tetrahedra. 
    {\bf e.} A comparison of the experimental data (upper panel) with the computed Raman response from the two-spinon continuum (lower panel).  Two-spinon  Raman  response is computed for the parallel channel (blue curve) for an arbitrary incoming light polarization in the $[111]$ plane [$\mathbf{e}_\mathrm{in}^{||}=\mathbf{e}_\mathrm{out}^{||}=(0.80,-0.26,-0.54)$  and  for the cross channel (red curve)  with $\mathbf{e}_\mathrm{in}^\perp=(0.80,-0.26,-0.54)$, $\mathbf{e}_\mathrm{out}^\perp=(-0.16,0.77,-0.61)$. The results are in a good agreement with the experimental data ($T = 20 $~K), where only the parallel channel gives a strong Raman intensity. {\bf f.} A comparison of the experimental data at $T= 7~K$ (upper panel) with the computed Raman response from the two-magnon excitation of all-in-all-out ordered Nd moments (M4).  Two-magnon Raman response computed for the  same light polarizations as in {\bf e}. In the experimental data M4 feature is weak compared to other observed excitations. 
    {\bf g.}    
    A summary of experimental Raman spectra of \nio\ in three different states: in paramagnetic semimetalic state at $T=35~K$, crystal field excitation (CEF) of Nd is observed at 25~meV, at $T= 16~K$, where M1 feature of Nd-spinon continuum is observed together with one-magnon excitations of Ir M2,M3; And in the state where both Nd and Ir moments are ordered all-in-all-out at $T=5~K$. Note that   M2 and M3 one-magnon excitations of Ir are found at the same frequencies as at $T=16~K$, while M4 two-magnon excitation of Nd order appears instead of the spinon continuum  (insets depict the 2I2O state and AIAO state of Nd moments, respectively). All three plots are showing the Raman response  in (x,x) channel.}
	\label{fig:spinon}
\end{figure*}

{\cred  Additionally, a comparison with well-studied spin dynamics in Nd$_2$Zr$_2$O$_7$ \cite{Petit2016, Benton2016, Lhotel2021} and Nd$_2$Hf$_2$O$_7$ \cite{Anand2015} reveals that Nd$^{3+}$ moment dynamics in Nd$_2$Ir$_2$O$_7$ manifest at significantly higher energy scales. This difference can be attributed to the unique magnetic properties of the B-site ions in Nd$_2$Ir$_2$O$_7$, as opposed to the non-magnetic B-site ions found in Nd$_2$Zr$_2$O$_7$ and Nd$_2$Hf$_2$O$_7$. These observations highlight the crucial interplay between spin dynamics in Nd$_2$Ir$_2$O$_7$.

Unfortunately, the origin of this magnetic continuum remains unresolved. Previous studies on the low-energy dynamics of Nd magnetic moments in \nio\ using inelastic neutron scattering~\cite{Tomiyasu2012} identified a distinct dispersionless mode around 1 meV, attributed to the splitting of the ground doublet of Nd$^{3+}$ ions in an AIAO state created by local Ir fields, estimated at around 1.3 meV. However, neutron scattering data at higher energies, particularly around 15 meV, is lacking.

We speculate that the continuum of excitations observed at 14 meV in the Raman response at intermediate temperatures may arise from fluctuations associated with the spin-ice phase of Nd$^{3+}$ moments. These fluctuations give rise to collective excitations, characterized by gapped and deconfined spinons, as well as emergent gapless gauge modes and emergent photons \cite{Ross2011,Hao2014,Fu2017}. However, while the low-energy Raman response of the quantum spin ice (QSI) phase can originate from both types of modes, the emergent photons, due to their low energy, fall below the detection threshold of our experimental setup. As a result, the Raman response in this temperature range is primarily dominated by spinon excitations, which manifest as a broad two-spinon continuum, specifically identified as the observed M1 mode.

To explain the formation of this spin-ice phase characterized by the 2-in/2-out (2I2O) spin configuration of Nd moments, we identify two plausible scenarios.
 The first scenario suggests that the Hamiltonian for Nd moments in Nd$_2$Ir$_2$O$_7$ mirrors that in other Nd pyrochlores, determined by the pyrochlore lattice symmetry and the dipole-octupolar nature of Nd moments. The key distinction, however, lies in the significantly renormalized coupling constants, enhanced by superexchange interactions mediated through extended Ir orbitals.
The local environment of  Nd magnetic moments,   schematically depicted in Fig.~\ref{fig:spinon} {\bf a-b}), shows how  partially filled extended 5d orbitals of Ir$^{4+}$ ions provide  many additional superexchange paths  between Nd$^{3+}$ ions. 
 
 The second scenario emphasizes the semimetallic nature of the low-temperature phase, where low-energy electronic excitations may influence the exchange interactions between Ir and Nd moments. This concept is partially illustrated in the Supplemental Information in Ref. \cite{Tian2016}, which suggests that weak Ir electron correlations, relative to the Kondo interactions between the Ir and Nd subsystems, favor 2I2O spin configurations for Nd moments over the AIAO order. This scenario provides an alternative explanation for the emergence of spin-ice behavior in the Nd subsystem, highlighting the influence of the semimetallic state and its electronic excitations.

 It is important to note that while there are significant differences between these two scenarios, they are not mutually exclusive. Instead, together, they contribute to the enhancement of Nd-Nd couplings and the spin-ice  dynamics of Nd magnetic moments.

Besides renormalizing the Nd-Nd interactions, below 33K ordered Ir$^{4+}$ moments provide a local field $h_{\rm loc}$ that acts on the Nd$^{3+}$ moments through the superexchange coupling $J_{\mathrm{Nd-Ir}}$.
This local field also contributes to the dynamics of the Nd$^{3+}$ moments\cite{Tomiyasu2012,Tian2016}. In the mean-field sense, this local field is proportional to
a net effective magnetic moment $\langle S_{\mathrm{net,Ir}}^z\rangle$ generated by six Ir moments $\langle m_{\rm Ir}\rangle$ neighboring the Nd ion that point along the local  $z$ direction (one of the global [111] axes) on each Nd site 
\cite{Chen2012,Kapon2022}
[see Fig.~\ref{fig:spinon} {\bf c}]. The net moment   acting on Nd ion  is equal to $\langle S_{\mathrm{net,Ir}}^z\rangle=2\langle m_{\rm Ir}\rangle/\mu_B$  because  four  out of six neighboring Ir moments  sum up to zero.

This local field induces a non-vanishing AIAO order on the Nd sublattice just below 33 K, once the Ir moments achieve their AIAO configuration. At temperatures near 33 K, the AIAO  order in Nd subsystem is small due to the smallness of $\langle S_{\mathrm{net,Ir}}^z\rangle$.  As a result, the system predominantly exhibits the 2I2O ice state, with only a small fraction showing AIAO order.

In the intermediate temperature range, $T^*_{\mathrm{Nd}}<T<T^{N}_{\mathrm{Ir}}$, $\langle S_{\mathrm{net,Ir}}^z\rangle$ increases, more and more Nd moments are driven into the AIAO state by the local field. As the temperature decreases further, particularly below 15 K, the prevalence of the 2I2O order diminishes, leading to the dominance of magnon excitations in the Raman response associated with the AIAO order of the Nd$^{3+}$ magnetic moments.

The competition between the 2I2O spin ice phase and AIAO order is further detailed in the Supplementary Information, where we analyze a minimal model incorporating the superexchange interactions favoring the 2I2O phase and the local field induced by the ordered Ir sublattice that supports AIAO order. Using the slave-particle formulation of Ref.\cite{Hao2014}, we derive the spinon excitation spectrum:
\begin{align}
\omega(\mathbf{k})=\frac{1}{2}\left(h_{\mathrm{loc}}\pm\tilde{J}_{z}\sqrt{1-\frac{\tilde{J}_{xy}}{2\tilde{J}_{z}}\sum_{\alpha\neq\beta}^{x,y,z}\cos\frac{k_\alpha}{2}\cos\frac{k_\beta}{2}}\right),
\end{align}
 where $\alpha,\beta={x,y,z}$ are the three global cubic directions.

}

The Raman response from the two-spinon excitation continuum in the parallel and cross polarizations shown in  the lower panel Fig.~\ref{fig:spinon}{e}~\footnote{the  incoming and outgoing light polarization  are the same as in the case for the one-magnon response shown in Fig. \ref{fig:magnon}{c} and { d}.} is computed using  the Raman operator for the QSI derived in Ref. \cite{Fu2017}.  $  \mathcal{R}_{\small{\mathrm{Nd}}}=\sum_{\langle ij\rangle}\left[(\mathbf{e}_{\mathrm{in}}\cdot\boldsymbol{\mu}_i)(\mathbf{e}_{\mathrm{out}}\cdot\boldsymbol{\mu}_i)+(\mathbf{e}_{\mathrm{in}}\cdot\boldsymbol{\mu}_j)(\mathbf{e}_{\mathrm{out}}\cdot\boldsymbol{\mu}_j)\right]{\mathcal{H}}_{\mathrm Nd}^{ij}$,
 where $\boldsymbol{\mu}_i$ and $\boldsymbol{\mu}_j$ denote the  relative position vectors of spinons associted with ${\mathcal{H}}_{\mathrm Nd}^{ij}$ (see Fig.~\ref{fig:spinon}{d}).
With the parameters set at $\tilde{J}_z=12.0$ meV, $\tilde{J}_{xy}=-3.0$ meV, and $h_{\mathrm{loc}}=12.0$ meV, our results for the two-spinon Raman response give good qualitative description of the M1 band which is  shown in  the upper panel of Fig.~\ref{fig:spinon}{e}, where only the parallel channel gives a strong Raman intensity.

{\cred We also computed the Raman response  from  Nd  moments in the AIAO phase below $T^*_{\mathrm{Nd}}$, where the M4 mode appears at  around 33 meV in  both polarizations [see Fig.\ref{fig:fig1}{ (b,c)} and Fig.\ref{fig:spinon}{g}]. 
From the energy of this mode we assume that it is two-magnon Raman response, and compute two-magnon  response for the AIAO state  of Nd ions \footnote{The absence of $\tilde{\tau}^z\tilde{\tau}^x$ and  $\tilde{\tau}^z\tilde{\tau}^y$ terms in the minimal model discussed in the Supplementary information rules out the one-magnon excitation origin of M4.}.}  We note that the computed intensity of the two-magnon response has stronger polarization dependence than the experimentally observed one. This might be related to the Loudon-Fleury approximation \cite{Fleury1968} we used in our calculations. 
A minimal remedy in our calculation is to introduce an anisotropy that makes  $\delta\equiv|\tilde{J}_x-\tilde{J}_y|\sim0.0002$ meV.
 Previously, it has been demonstrated that in systems featuring multiple superexchange paths, the non-Loudon-Fleury terms may give a substantial contribution in the cross-polarization channel \cite{Yang2021}, {\cred which might lead to significant modification of polarization dependence in Fig.~\ref{fig:spinon} \textbf{e} and \textbf{f}. } However,  considering these terms in this case  is  a challenging task that lies beyond the scope of this paper.

{\it Splitting of 42 meV E$_g$ phonon.}-- Lastly, we would like to address the distinctive aspects of phonon dynamics observed in our  Raman data. Below T$^{Ir}_N$, the changes in the phonon spectrum become evident, most notably in the E$_g$ phonon at around 42 meV. This phonon broadens and splits, with spectral weight shifting towards the higher frequency component (see Fig.~\ref{fig:fig1}({b,g})). These observations provide insights into how the lattice of \nio\ responds to magnetic ordering 
 and changes  in the electronic structure.  While a separate study is required to  fully understand the lattice response in such a complex system as \nio\, 
  featuring two interacting magnetic sublattices and a complex electronic structure, here we provide
  some qualitative   description.
  
  The observed lattice response can  stem from two main scenarios: (i) Magneto-elastic coupling and (ii) Electron-phonon coupling. In the first scenario, the magneto-elastic coupling arises from the dependence of  the   coupling between magnetic moments on  the distance between them. It is usually large in the systems with strong spin-orbit coupling. In the paramagnetic phase, the $E_g$ phonon couples to the linear combinations of magnetic  bond energies transforming under the compatible 2D irrep of the paramagnetic group. Below the AIAO ordering transition, since  phonons and magnetic fluctuations are linearly coupled by magneto-elastic coupling, the system's symmetry reduces to $m\bar{3}m'$, in which 2D irrep $E_g$ splits into two distinct 1D irreps. Subsequently, one of the two phonons couples with the compatible one-magnon mode, M2, resulting in the splitting of the $E_g$ phonon mode of the paramagnetic phase  below $T_{\mathrm{Ir}}^N$. It is essential to note that this phenomenon is solely dynamical and doesn't necessarily induce lattice distortion. Further details regarding the symmetry analysis can be found in Section S2 of the SI.

In the second scenario, the splitting of the $E_g$ phonon modes is attributed to electron-phonon coupling. Although we do not believe this scenario applies to \nio, previous investigations in other iridium pyrochlores, like Pr$_2$Ir$_2$O$_2$~\cite{Xu2022phonons} and Eu$_2$Ir$_2$O$_2$~\cite{Ueda2019}, have indicated that electron-phonon coupling can significantly impact phonon dynamics.

\section{Conclusion}
 
{\cred 
The complex temperature dependence of the Raman response in the magnetic pyrochlore \nio\ reveals the intricate interplay between electronic, magnetic, and lattice degrees of freedom. Below $T_\mathrm{N}^{\small \rm{Ir}}$, we observe AIAO ordering of the Ir moments, accompanied by $\Gamma$-point magnons. Our calculations, fitted to the observed magnon frequencies, provide key parameters for the magnetic interactions between Ir$^{4+}$ moments.

The strong magnetic coupling between the Ir and Nd sublattices significantly enhances the interactions among Nd$^{3+}$ moments, likely driving 2I2O fluctuations and producing a broad Raman feature centered at 14 meV in the intermediate temperature range ($20~\mathrm{K} < T < T^{N}_{\mathrm{Ir}}$). We speculate that this broad mode could represent a collective excitation from the spinon continuum of the intermediate spin-ice phase of Nd$^{3+}$ moments. The AIAO ordering of Nd$^{3+}$ moments also sets in at a relatively high temperature. This enhancement of Nd-Nd interactions makes \nio\ an intriguing platform for exploring rare-earth magnetism and  potentially quantum spin ice  physics.

Finally, we emphasize that the AIAO ordering of Ir moments coincides with the Luttinger semimetal-to-Weyl semimetal transition in the electronic system
 \cite{Nikolic2022}. While the exchange couplings between Nd moments are mediated by Ir ions, they do not directly depend on the Ir magnetic order. The absence of 2I2O fluctuations above 33 K, along with the emergence of the M1 peak only below this temperature, can be attributed to the dominance of electronic excitations in the low-energy spectrum, which effectively mask 2I2O behavior. Additionally, the magnetic transitions in the Nd subsystem occur as a crossover, rather than a sharp phase transition, allowing for the coexistence of AIAO and 2I2O states down to low temperatures. Consequently, the M1 peak  potentially associated with the 2I2O phase gradually diminishes, while the M4 peak, linked to two-magnon excitations of the Nd  AIAO , becomes more pronounced as the temperature drops.}

\section{Methods}
\nio~single crystals were grown by the KF-flux method~\cite{Millican2007} and possess  as-grown
octahedron-shaped (111) facets. Raman scattering spectra were collected from (111) cleaved surface of a single crystal of \nio\ using two different Raman setups. The first setup allowed us to measure non-polarized Raman  spectra in the temperature range from 300 to 5 K and in the spectral range down to 10 meV. Selected spectra from these measurements are shown in Fig.3g.  The measurements were done using Horiba Labram HR Evolution spectrometer equipped with Olympus microscope and an ultra narrow notch filter.  The spectra were excited using a 532 nm laser radiation. Sample was placed in a He flux cryovac micro Konti cryostat. The exact temperature of the sample was obtain by a comparison of Stokes and anti-Stokes intensities. 

The second Raman scattering apparatus allowed us to obtain polarized spectra in the temperature range 300-15 K for frequencies down to 3.5 meV (triple monochromator option) and at temperatures down to 7~K with the spectral range limited by 12 meV at low frequencies (single monochromator and edge filter option). Selected spectra are shown in Fig. 1 b-c. These measurements were done using the Jobin-Yvon T64000 triple monochromator spectrometer equipped with a liquid nitrogen cooled CCD detector with spectral resolution of 2~cm$^{-1}$. 514.5 nm line of Ar$^+$-Kr$^+$ mixed gas laser was used as the excitation light. The intensity of the incident light was 3 mW at 4 K and 10 meV above 4 K for single monochromator, and 15 meV for triple monochromator before the crystal and the laser heating was estimated to be about 1 K per 1mW.  The measurements were performed in  pseudo-Brewster's  geometry using an elliptically shaped  laser probe of 50 by 100 microns in size. The polarization-resolved spectra were measured in four configurations: $\hat{z}(xx)z$, $\hat{z}(xy)z$, $\hat{z}(RR)z$, and $\hat{z}(RL)z$, where $R$($L$) denotes the right (left) circular polarization, which allow to detect scattering channels of symmetries summarized in Table~\ref{table:intensity}.   For low temperature measurements the sample was  mounted on the cold-finger of Janis ST-500 cryostat, which can be cooled down to 4 K without laser heating.  The presented Raman response $\chi''(\omega, T)$ was normalized on the Bose-Einstein factor $[n(\omega, T)+1]$, where $n(\omega, T) = [\exp(\hbar \omega / k_\mathrm{B}T) - 1]^{-1}$ is the Bose occupation factor.

To correct for  the small deviations  of the intensity of Raman response in different measurements, which were less than 20\%  for the same excitation power, all the spectra were normalized to the intensity of the $A_{1g}$ phonon at 63 meV and the band at 82 meV.

\begin{table}[htb]	
	\centering
	\caption{Components of Raman tensor for $(x,x)$ and $(x,y)$ polarizations in measurements geometry when polarization $x$ and $y$ of electrical vectors $e_{in}$ and $e_{out}$ are parallel to the   [111] crystalographic plane}
        \label{table:intensity}
	\begin{ruledtabular}
	\begin{tabular}{cccc}
		Geometry & $A_{1g}$ & $E_{g}$ & $T_{2g}$\\
		\hline
		$(x,x)$ & $a^2$ & $b^2$ & $d^2$\\
		$(x,y)$ & 0 & $b^2$ & $\frac{2}{3}d^2$\\
		$(R,R)$ & 0 & $2b^2$ & $\frac{4}{3}d^2$\\
		$(R,L)$ & $a^2$ & 0 & $\frac{1}{3}d^2$
	\end{tabular}
	\end{ruledtabular}
\end{table}

Fitting of the experimental spectra was done  by means of least squares assuming Lorentzian peak shapes for magnetic excitations and phonons, with the resulting function 
\begin{equation}
    \chi''(\omega) = \chi''_0(\omega) +  \frac{1}{2\pi}\sum_{i = 1}^N  \frac{A_i\Gamma_i}{(\omega - \omega_i)^2 + (\Gamma_i / 2)^2},
\end{equation}
where $\omega_i$, $\Gamma_i$, and $A_i$ correspond to the center, full width, and amplitude of the $i$th Lorentzian peak, and $\chi''_0(\omega)$ is the continuous electronic background. The detailed fitting results are shown in the supplement information.

\section{Acknowledgements}
The authors are thankful to  C.~Broholm  for useful discussions. This work was supported as part of the Institute for Quantum Matter, an Energy Frontier Research Center funded by the U.S. Department of Energy, Office of Science, Basic Energy Sciences under Award No.~DE-SC0019331. This work in Japan is partially supported by CREST (Grant Number: JPMJCR18T3 and JPMJCR15Q5), by New Energy and Industrial Technology Development Organization (NEDO), by Grants-in-Aids for Scientific Research on Innovative Areas (Grant Number: 15H05882 and 15H05883) from the Ministry of Education, Culture, Sports, Science, and Technology of Japan, and by Grants-in-Aid for Scientific Research (Grant Number: 19H00650). The work of Y.Y. and N.B.P. was supported by the U.S. Department of Energy, Office of Science, Basic Energy Sciences under Award No. DE-SC0018056.
N.B.P.  also acknowledges the hospitality and partial support  of the Technical University of Munich – Institute for Advanced Study and  the Alexander von Humboldt Foundation.

\section{Author contributions}
 N.~D. conceived the idea of the experiment. T.~O. and Y.~Q., and S.~N. grew the crystals. Y.~X., J.~T., and N.~D. collected and analyzed the Raman scattering data. Y.~Y.and N.~P. calculated the magnon dispersion and two-spinon excitations. N.~D, D.v.d.M., Y.~Y.,Y.~X., and N.~P. wrote the manuscript.

\section{Competing interests}
The authors declare no competing interests.

\bibliography{./PyrochloresNIO}

\newpage

\onecolumngrid
\begin{center}
 {\bf \large Supplement Information for the manuscript ``Ramification  of complex magnetism  in Nd$_2$Ir$_2$O$_7$ observed by Raman scattering spectroscopy''}    
\end{center}
\twocolumngrid

\author{Yuanyuan Xu}
\affiliation{Institute for Quantum Matter and Department of Physics and Astronomy, Johns Hopkins University, Baltimore, Maryland 21218, USA}

\author{Yang Yang}
\affiliation{School of Physics and Astronomy, University of Minnessota, Minneapolis, MN 55455, USA}

\author{J\'er\'emie Teyssier}
\affiliation{Department of Quantum Matter Physics, University of Geneva, Geneva, Switzerland}

\author{Takumi Ohtsuki}
\affiliation{Institute for Solid State Physics, University of Tokyo, Kashiwa, Chiba 277-8581, Japan}

\author{Yang Qiu}
\affiliation{Institute for Solid State Physics, University of Tokyo, Kashiwa, Chiba 277-8581, Japan}

\author{Satoru Nakatsuji}
\affiliation{Institute for Quantum Matter and Department of Physics and Astronomy, Johns Hopkins University, Baltimore, Maryland 21218, USA}
\affiliation{Institute for Solid State Physics, University of Tokyo, Kashiwa, Chiba 277-8581, Japan}
\affiliation{Department of Physics, University of Tokyo, Bunkyo-ku, Tokyo 113-0033, Japan}
\affiliation{Trans-scale Quantum Science Institute, University of Tokyo, Bunkyo-ku, Tokyo 113-0033, Japan}
\affiliation{CREST, Japan Science and Technology Agency, Kawaguchi, Saitama 332-0012, Japan}
\author{Dirk van der Marel}
\affiliation{Department of Quantum Matter Physics, University of Geneva, Geneva, Switzerland}

\author{Natalia  B. Perkins}
\affiliation{School of Physics and Astronomy, University of Minnessota, Minneapolis, MN 55455, USA}

\author{Natalia Drichko}
\affiliation{Institute for Quantum Matter and Department of Physics and Astronomy, Johns Hopkins University, Baltimore, Maryland 21218, USA}
\email{Corresponding author. Email:drichko@jhu.edu}

%\date{\today}

%\maketitle
In this Supplementary Information, we present details of our analysis of the Raman scattering data discussed in the main text, along with additional data used to explain and support the analysis. Specifically,
in Section S1.1, we compare the scattering data in (x,x) and (x,y) polarization channels; in Section S1.2, we provide additional information used for the phonon's symmetry assignment;
in Section S1.3, we analyze the crystal field excitations of Nd$^{3+}$ ions.
In Sections S2, we present the symmetry analysis of M2 and M3 one-magnon excitations of Ir magnetic moments.
  In Sections S3,
 we provide details on the computation of the two-spinon Raman responses from the Nd magnetic subsystem.

\section{
%Experimental Raman scattering spectra. 
Additional information  for the Raman scattering data}

\subsection{Comparison of  the (x,x) and (x,y) scattering channels}

\begin{figure*}
	\centering
	\includegraphics[width=\linewidth]{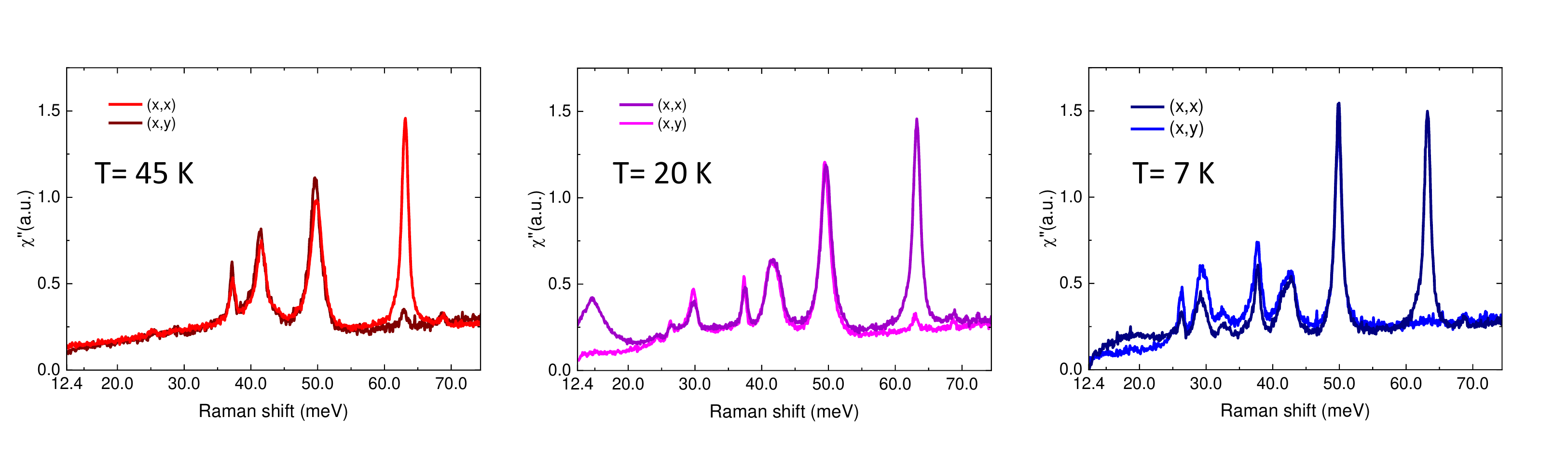}
    \vspace*{-1cm}
	\caption{ A comparison of (x,x) and (x,y) scattering channels for  Nd$_2$Ir$_2$O$_7$ at 
45 K (above $T_\mathrm{N}^{\small\rm{Ir}}$),
20 K (below  $T_\mathrm{N}^{\small\rm{Ir}}$ but above $T^\mathrm{*}_{\small\rm{Nd}}$), and  7 K (below $T^\mathrm{*}_{\small\rm{Nd}}$).
%  $ > T_\mathrm{N}^{\small\rm{Ir}}$, $T_N^{Ir}  > $T=20 K$ >T^\mathrm{*}_{\small\rm{Nd}}$, and 7 K $ <  T^\mathrm{*}_{\small\rm{Nd}}$.
}
 \label{fig:Polarizations_wide}
\end{figure*}

%As we discussed in the Methods section, 
Raman scattering was measured from (111) cleaved surface of the crystal of \nio\ in both (x,x) and (x,y) polarization channels.
Fig.~\ref{fig:Polarizations_wide} presents a comparative analysis of these two scattering channels at temperatures 45 K, 20 K, and 7 K.
Note the absence of A${1g}$-symmetry excitations, the spinon continuum, and the phonon at approximately 50 meV in the (x,y) scattering channel.

\begin{figure*}
	\centering
	\includegraphics[width=\linewidth]{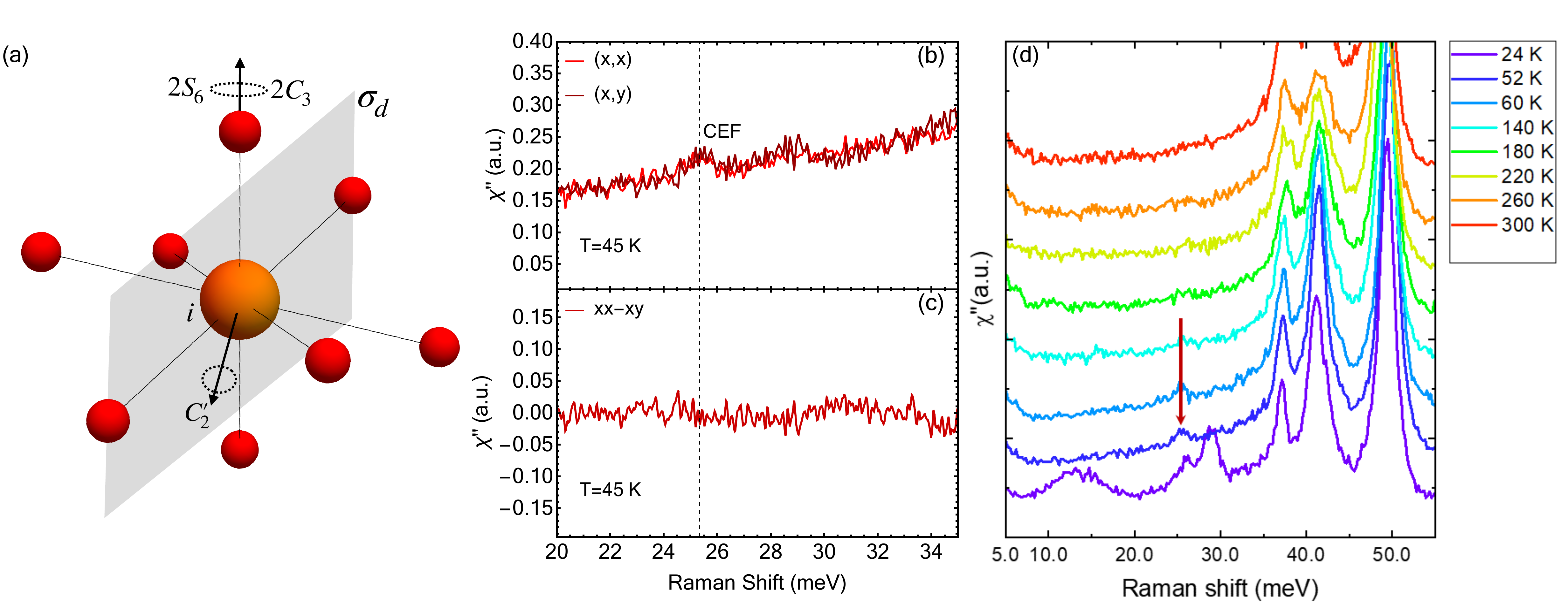}
	\caption{(a) Local oxygen ion environment of a Nd ion described by $D_{3d}$ point group. (b) A zoom-in plot of Raman response shown in Fig.\ref{fig:Polarizations_wide} between $20$ meV and $35$ meV for \nio\ at $T=45$ K (c) xx-xy curve eliminates $E_g$ channel, and shows no  signature of the CEF excitation, confirming the symmetry selection rule of the CEF excitation. (d) Temperature dependence of Raman scattering of Nd$_2$Ir$_2$O$_7$. We point on the weak excitation at 25~meV with a red arrow. We assign this excitation to the lowest crystal field excitation. It is observed in the spectra above $T^\mathrm{*}_{\small\rm{Nd}}$ and overlaps with a magnon excitations at lower temperatures. }
 \label{fig:SI_CEF}
\end{figure*}

\subsection{Phonon's symmetry assignment}

\begin{table}[!htb]
    \caption{Frequencies of the Raman active phonons.}
    \label{table:phonon}
    \begin{tabular}{|c|c|}
            \hline
            Mode & Frequency (meV) \\
            \hline
            $T_{2g}^{(1)}$ & 37.4 \\
            $E_g^{(1)}$ & 41.4  \\
            $E_g^{(2)}$ & 42.6 \\
            $T_{2g}^{(2)}$ & 49.6 \\
            $A_{1g}$ & 63.1 \\
            $T_{2g}^{(3)}$ & 68.6 \\
            $T_{2g}^{(4)}$ & 76.8 \\\hline
    \end{tabular}
\end{table}

\begin{figure}%[!htb]
  %  \hspace*{-0.5cm}
    \centering
    \includegraphics[width=1\linewidth]{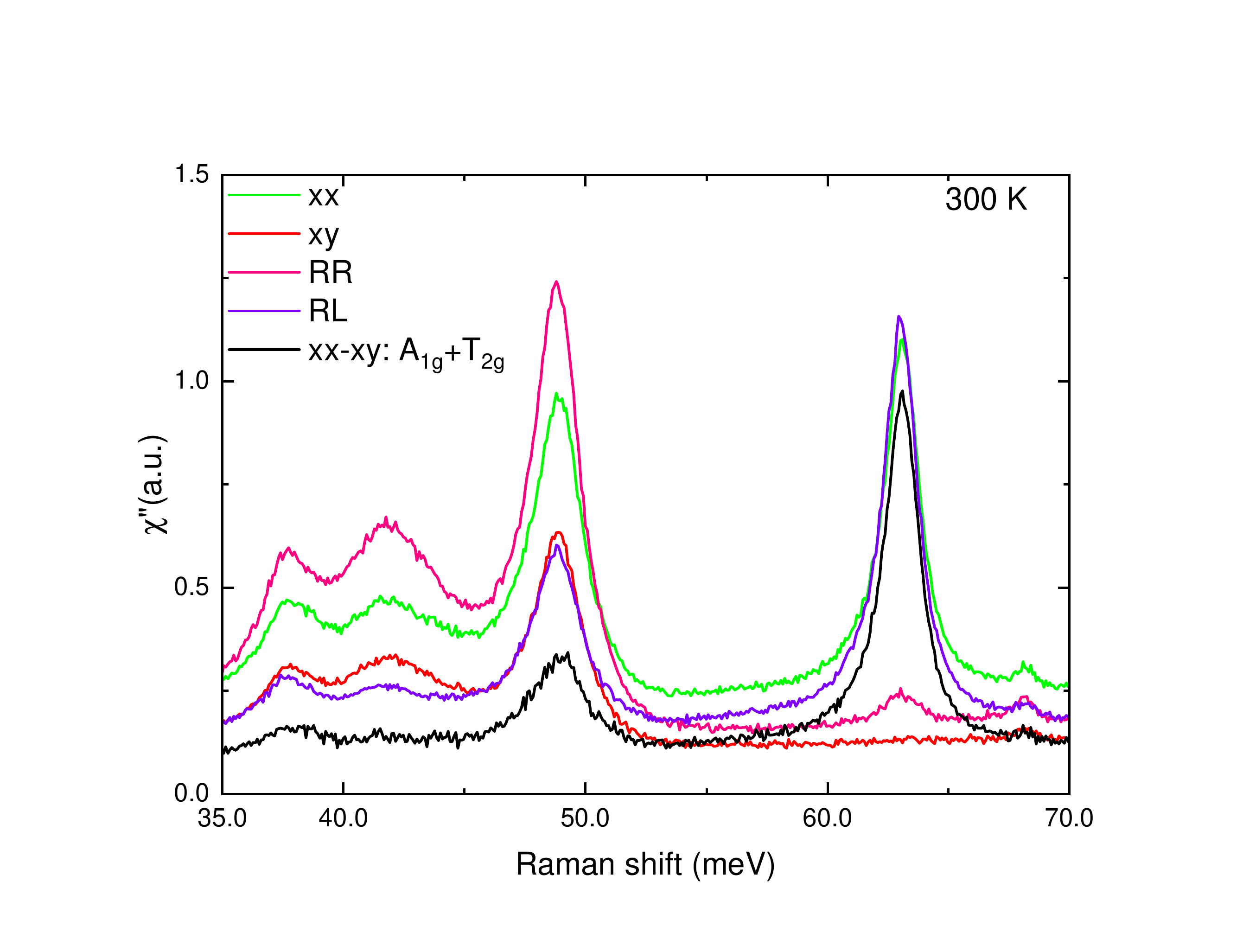}
   % \vspace*{-1cm}
    \caption{Polarization dependence of Raman scattering  in phonons of Nd$_2$Ir$_2$O$_7$ at 300 K. 
     Both RR and RL scattering exhibit a minor polarization leakage effect.
    The  xx-xy curve excludes E$_g$ channel and demonstrates that the excitation at 42 meV belongs to E$_g$ scattering channel, confirming the phonon assignment.}
    \label{fig:Polarizations}
\end{figure}

The pyrochlore lattice has $Fd\bar{3}m$ (No.~227) space group corresponding to $O_h$ ($m\bar{3}m$) point group. By symmetry, \nio\ has 6 Raman-active phonons, found at frequencies close to that observed in other pyrochlore iridate materials~\cite{Hasegawa2010,Xu2022phonons,Ueda2019,Nguyen2021}. Polarization dependence of the phonons intensity at 300 K is presented in Fig.~\ref{fig:Polarizations}. 
To establish  the phonons assignment  summarized in Table.~\ref{table:phonon},  we used the experimental data presented  in Figure 1 of the main text and Fig.~\ref{fig:Polarizations}, as well as   previously published experimental spectra and phonon density functional theory (DFT) calculations for Eu$_2$Ir$_2$O$_2$, Pr$_2$Ir$_2$O$_2$, and Y$_2$Ir$_2$O$_2$~\cite{Xu2022phonons,Ueda2019,Nguyen2021}. The spectra of pyrochlore iridates presented in these publications are in agreement with the first publication on Raman scattering of pyrochlore iridates, Ref.~\cite{Hasegawa2010}, while the phonon assignment was revised in the later manuscripts, based on the DFT calculations. 
 We note that only the oxygen atoms O,  bonded to two Nd$^{3+}$ and two Ir$^{4+}$, and O$'$, tetrahedrally bonded to four Ir$^{4+}$, are Raman active.   The symmetry of the  6 phonons is determined experimentally by the polarization-resolved measurements according to the polarization selection rule, as described in the section Methods of the main text.

\subsection{Crystal field excitations of Nd$^{3+}$}

In Nd$_2$Ir$_2$O$_7$, each Nd ion locally has D$_{3d}$ point group symmetry due to surrounding oxygen ions [shown in Fig.~\ref{fig:SI_CEF} (a)]. As a result, J=9/2 multiplet will be split into 5 doublets due to the dominant crystal electric field (CEF) component $B_2^0 O^0_2$ \cite{Gardner2010}. So far the estimates for these energy level splitting have relied on the previous neutron scattering results~\cite{Watahiki2011,Xu2015}. Raman scattering spectroscopy, with its enhanced energy resolution, allows to significantly enhance the precision of these measurements.
In Raman scattering spectra,   %E$_g  \rightarrow$ E$_g$
the $\Gamma^+_{56}\rightarrow\Gamma^+_{4}$ ($E_{3/2g}\rightarrow E_{1/2g}$ in Mulliken notation \cite{Inui1990}) transition  around $25$ meV (in agreement with \cite{Watahiki2011,Xu2015}) is observed in both parallel (x,x) and cross-polarized (x,y) scattering channel.
The selection rules governing the allowed symmetry of the CF excitation from one doublet to another are determined by the direct product of their respective representations \cite{Kiel1969}. In this case, the direct product of the representations can be found as:  
$\Gamma^+_{56} \bigotimes\Gamma^+_{4}=\Gamma_3^++\Gamma_3^+$ ($E_{3/2g} \bigotimes E_{1/2g}=E_g+E_g$ in Mulliken notation). Therefore, the observed CEF excitation is expcected to have the same polarization dependence as $E_g$ phonon [Shown in Fig.~\ref{fig:SI_CEF} (b) and (c)].

Typically, CEF excitations are at least one order of magnitude weaker than phonons, and appear in the spectra only at low temperatures. We clearly observe a CEF of Nd$^{3+}$ at 25 meV at temperatures above the ordering temperature of Ir$^{4+}$ moments T$_N^{Ir}$,  as shown  at  T=35 K in the upper panel of Fig. 3 {\bf g} of the main text.  In Fig.~\ref{fig:SI_CEF} (d),  we present the temperature dependence of the spectra of Nd$_2$Ir$_2$O$_7$.
This figure clearly shows that the CEF excitation  becomes visible in the spectra  at temperatures below  100 K. 
Notably, it  is very close in energy  to the  Ir magnon band M2, which emerges below  T$_N^{Ir}$= 33 K. As a result, distinguishing these two excitations may require more high-resolution measurements.

\section{Symmetry analysis of the one-magnon excitation}
 In this section, we provide the symmetry analysis of the  one-magnon  M2 and M3 modes appearing in the Raman spectra below $T^\mathrm{N}_{\small\rm{Ir}}$.
%One-magnon 
These excitations come from the $\mathbf{k}=0$ spin-wave excitations of the Ir-Ir interacting Hamiltonian:
%given in the main text
\begin{align}
    \mathcal{H}_{\small{\mathrm{Ir}-\mathrm{Ir}}}
    =&\sum_{\langle ij\rangle_\nu}\Bigl[ J\,\mathbf{S}_i\cdot\mathbf{S}_j+K\, S_i^{\alpha_\nu}S_j^{\alpha_\nu}\nonumber\\&+\sigma_\nu\Gamma_{ij}\,(S_i^{\beta_\nu} S_j^{\gamma_\nu}+ S_i^{\gamma_\nu} S_j^{\beta_\nu})+\mathbf{D}_{ij}\cdot(\mathbf{S}_i\times\mathbf{S}_j)\Bigr],
    \label{eqn:Ir-Ir}
\end{align}
 introduced and discussed in the main text.
With the linear spin wave analysis on the AIAO state, we obtain two $\mathbf{k}=0$ modes, $\epsilon_1$ (non-degenerate) and $\epsilon_2$ (three-fold degenerate).
 The best comparison with the experimental data is obtained with $(J,K,\Gamma,D)=(6.1,\,-5.4,\,3.0,\,4.1)\,\text{ meV}$. 

\begin{figure}[t]
    \centering
    \includegraphics[width=0.4\textwidth]{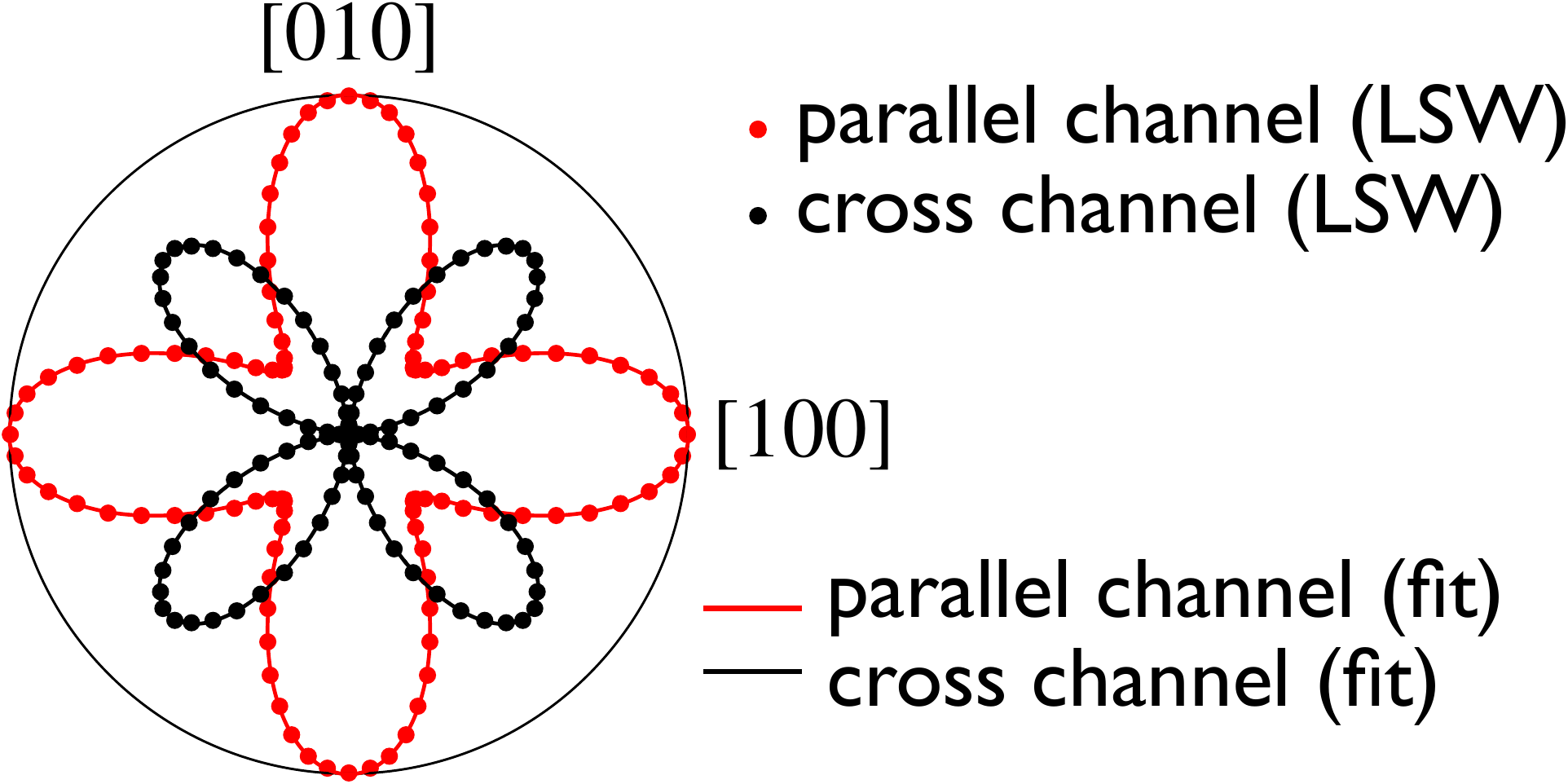}
    \caption{Comparison of the angular dependence of Raman response from the M2 one-magnon mode obtained in the linear spin wave (LSW) theory with the one obtained from the magnetic Raman tensor analysis.}
    \label{fig:fit_magnons}
\end{figure}

The magnetic point group of the AIAO state is 
$m\bar{3}m'$, formed by the unitary point group $T_h$ and the non-unitary point group $\Theta\sigma_d T_h$ \cite{Cracknell1966,Cracknell1969}.
There is only one Raman-active three dimensional (3D) irreducible corepresentation of $m\bar{3}m'$. Thus, the three-fold degenerate   mode  $\epsilon_2$ and, correspondingly, M3 peak transform as
$D\Gamma_4$($DT_g$) irreducible corepresentation \cite{Cracknell1969}. 
The $\epsilon_1$ mode is not degenerate, so it should correspond to one of the one-dimensional (1D) irreducible corepresentation.
There are three 1D irreducible corepresentations of $m\bar{3}m'$, so we need to determine which irreducible corepresentation the state corresponds to. We first fit the angular dependence of the Raman response shown in Fig.~\ref{fig:fit_magnons}
with the magnetic Raman tensors \cite{Cracknell1969}
\begin{align}
    \begin{pmatrix}
        \omega & 0 & 0\\
        0 & \omega^* &0\\
        0 & 0 & 1
    \end{pmatrix} \text{or } 
    \begin{pmatrix}
        \omega^* & 0 & 0\\
        0 & \omega &0\\
        0 & 0 & 1
    \end{pmatrix}
\end{align}
where $\omega=\exp(2\pi i/3)$. These two magnetic Raman tensors belong to two 1D irreducible corepresentations $D\Gamma_2$ and $D\Gamma_3$, originated from the $2D$ irreducible representation $E_g$ of the $T_h$ point group. 
  Applying $C_3$ symmetry to the eigenvector of the  
 $\epsilon_1$ mode, $C_3|\epsilon_1\rangle\rightarrow e^{2\pi i/3}|\epsilon_1\rangle$, confirms that it transforms under the irreducible corepresentation of $D\Gamma_2$. Thus,  magnetic M2 peak  belongs to
 the $D\Gamma_2$ irreducible representation of the magnetic point group $m\bar{3}m'$.

\section{ Magnetic response  from Nd  sublattice}
{\cred  In this section, we compute the M1 peak in the Raman spectra based on the premise that the spin-ice phase in Nd magnetic moments at intermediate temperatures emerges from the significant renormalization of superexchange interactions due to the influence of extended Ir orbitals.  This assumption is crucial, as the typical energy scale of superexchange interactions between Nd moments, originating from their localized f-orbitals, is generally  estimated as 0.01-0.1 meV in materials like Nd$_2$Zr$_2$O$_7$ and Nd$_2$Hf$_2$O$_7$~\cite{Anand2015,Xu2019}. This range is too small to account for the M1 excitation observed in Nd$_2$Ir$_2$O$_7$ around 14 meV within the 15-33 K temperature range. 

We propose that, in addition to conventional superexchange interactions involving direct exchange and oxygen ions, additional pathways facilitated by the extended, partially filled 5d orbitals of Ir$^{4+}$ ions enhance these interactions. The local environment of the Nd magnetic moments, as depicted in Fig. 3 {\bf a-b} of the main text, suggests that the 5d orbitals of Ir$^{4+}$ provide multiple superexchange routes between Nd$^{3+}$ ions. Due to their delocalized nature, these 5d orbitals can significantly renormalize the coupling terms in Eq.~(\ref{eq:xyz}), leading to stronger interactions than those typically associated with f-orbitals. 
Although a detailed microscopic analysis of these pathways using density functional theory (DFT) is beyond the scope of this paper, we assume  here that they both enhance couplings between Nd$^{3+}$ moments and support QSI 2I2O fluctuations.
}

\subsection{ Effective Hamiltonian }

{\cred 
We consider a 
minimal, symmetry-allowed nearest-neighbor exchange Hamiltonian for Nd${}^{3+}$ moments on the pyrochlore lattice in the presence of local field excerted  on Nd${}^{3+}$ by ordered Ir${}^{4+}$ into the AIAO order.

 The superexchange Hamiltonian for  dipolar-octupolar (DO) doublet degrees of freedom of Nd${}^{3+}$ proposed by Huang {\it et al } \cite{Huang2014} reads: 
 \begin{align}
    \mathcal{H}_{\mathrm{Nd-Nd}}=\sum_{\langle ij \rangle}&J_{x}\tau^x_i\tau^x_j+
    J_{y}\tau^y_i\tau^y_j+J_{z}\tau^z_i\tau^z_j\nonumber\\&+J_{xz}(\tau^x_i\tau^z_j+\tau^z_i\tau^x_j),
    \label{eq:spin_ice}
\end{align}
{\bf where the pseudospin-1/2 operators $\tau$ describe the ground state doublets of $J=9/2$ multiplet on  Nd${}^{3+}$ ion
in the local reference frame under the CEF.} The $\tau_x$ and $\tau_z$ operators have a dipolar character, while the $\tau_y$ operator is purely octupolar. This distinction ensures that the symmetry-allowed coupling terms exclusively involve operators of the same character.

 Next, it is customary to perform a local rotation around the local $y$ axis \cite{Huang2014,Benton2016,Xu2019} to eliminate $\tau_x-\tau_z$ coupling such that Eq. (\ref{eq:spin_ice}) becomes 
\begin{align}
    \mathcal{H}_{\mathrm{Nd-Nd}}=\sum_{\langle ij \rangle}&\tilde{J}_{x}\tilde{\tau}^x_i\tilde{\tau}^x_j+\tilde{J}_{y}\tilde{\tau}^y_i\tilde{\tau}^y_j+\tilde{J}_{z}\tilde{\tau}^z_i\tilde{\tau}^z_j,
    \label{eq:xyz}
\end{align}
with the rotation angle $\theta=\arctan[2J_{xz}/(J_x-J_z)]/2$ \cite{Benton2016}.  
This simple pyrochlore XYZ model (\ref{eq:xyz}) supports  both the QSI  2I2O phase  ($\tilde{J}_{z}>0$) and magnetically ordered AIAO order ($\tilde{J}_{z}<0$).

Another contribution to the dynamics of Nd${}^{3+}$ moments arises from a local field, $h_{\rm loc}$, exerted by ordered Ir$^{4+}$ moments ~\cite{Tomiyasu2012,Tian2016}.
In a mean-field approximation, this local field is proportional to the net effective magnetic moment $\langle S_{\mathrm{net,Ir}}^z \rangle$, which is generated by the six neighboring Ir moments, $\langle m_{\rm Ir} \rangle$, that surround each Nd ion and align along the local $z$ direction (one of the global [111] axes) at each Nd site \cite{Chen2012,Kapon2022} [see Fig.3 {\bf c} in the main text]. The net moment acting on the Nd ion is $\langle S_{\mathrm{net,Ir}}^z \rangle = 2\langle m_{\rm Ir} \rangle / \mu_B$, as four out of the six neighboring Ir moments cancel each other out.
 Thus, the direct
interaction between the Nd moment and the ordered Ir moment can be described as
\begin{align}
    \mathcal{H}_{\rm Nd-Ir}=-h_{\rm loc}\sum_i \tau_i^z\approx -h_{\rm loc}\sum_i \tilde{\tau}_i^z,
    \label{eqn:Nd-Ir}
\end{align}
where the local field $h_{\rm loc}=J_{\rm Nd-Ir}\langle S_{\mathrm{net,Ir}}^z\rangle$ 
\footnote{Note that this interaction assumes that Ir moments are localized and, therefore, differs both from effective f-d interaction describing coupling between itinerant 5d electrons of Ir and Ising-like Nd moments considered \cite{Chen2012,Kapon2022} and from the Kondo coupling of Nd and Ir spins exploited in \cite{Tian2016}.  In fact, such local field interaction has been proposed for Ho${}_2$Ir${}_2$O${}_7$ in \cite{Lefrancois2017} to explain magnetic fragmentation.} and the difference between the original local coordinate and the rotated coordinate is ignored.

Thus, to compute the two-spinon Raman responses, we  will use the following  Hamiltonian for Nd${}^{3+}$ magnetic moments 
\begin{align}
    \mathcal{H}_{\mathrm{Nd}}=&\mathcal{H}_{\mathrm{Nd-Nd}}+\mathcal{H}_{\mathrm{Nd-Ir}}\nonumber\\=&\sum_{\langle ij \rangle}\tilde{J}_{x}\tilde{\tau}^x_i\tilde{\tau}^x_j+
    \tilde{J}_{y}\tilde{\tau}^y_i\tilde{\tau}^y_j+ \tilde{J}_{z}\tilde{\tau}^z_i\tilde{\tau}^z_j-h_{\rm loc}\sum_i \tilde{\tau}_i^z,
    \label{eqn:renormalized}
\end{align}

\subsection{ Dynamics of Nd moments}

As discussed in the main text, the AIAO ordering of Ir magnetic moments coincides with the Luttinger semimetal-Weyl semimetal transition \cite{Nikolic2022} in the electronic system. It is important to note that while the exchange couplings between Nd moments are mediated by Ir ions, they are not directly dependent on Ir magnetic ordering. The absence of 2I2O fluctuations above 33 K, along with the emergence of the M1 peak only below this temperature, can be attributed to the dominance of electronic excitations in the low-energy spectrum above $T^{N}_{\mathrm{Ir}}$, which effectively mask the distinctive behavior of the 2I2O states.

To compute the spectrum of spinon excitations in the renormalized Hamiltonian (\ref{eqn:renormalized}), we use the slave-particle formalism from Ref.~\cite{Hao2014}. This formalism incorporates charge degrees of freedom,  $Q_\mathbf{x}$, which represent violations of ice rules on the A and B sublattices of the diamond lattice of tetrahedra. The spinon operators  $\psi_\mathbf{x}$ are introduced to express excitations, where the pseudospin operators $s_{\mathbf{x}\mu}^{\pm}$ are related to the vector gauge field $\mathbf{A}_{\mathbf{x}\mu}$, describing the interactions between spinons. In this approach we have 
\begin{align}
    Q_\mathbf{x}=\begin{cases}\sum_\mu \tilde{\tau}_{\mathbf{x},\mu}^z,\quad &\mathbf{x}\in\langle A\rangle,\\\sum_\mu -\tilde{\tau}_{\mathbf{x}-\hat{\boldsymbol{\mu}},\mu}^z,\quad &\mathbf{x}\in\langle B\rangle\end{cases}
\end{align}
 where $\mathbf{x} \in$ A or B sublattices of the diamond lattice of tetrahedra, and 
\begin{align}
    \tilde{\tau}_{\mathbf{x},\mu}^+&\equiv\tilde{\tau}_{\mathbf{x},\mu}^x+i \tilde{\tau}_{\mathbf{x},\mu}^y=\psi_\mathbf{x}^\dagger s_{\mathbf{x}\mu}^+\psi_{\mathbf{x}+\hat{\boldsymbol{\mu}}},\label{eqn:spinon-p}\\
    \tilde{\tau}_{\mathbf{x},\mu}^-&\equiv\tilde{\tau}_{\mathbf{x},\mu}^x-i \tilde{\tau}_{\mathbf{x},\mu}^y=\psi_{\mathbf{x}+\hat{\boldsymbol{\mu}}}^\dagger s_{\mathbf{x}\mu}^-\psi_\mathbf{x},\label{eqn:spinon-m}
\end{align}
where   $s_{\mathbf{x}\mu}^\pm$   denote the pseudospin operators, which can be further 
mapped to $\exp(\pm i\mathbf{A}_{\mathbf{x}\mu})$   with  $\mathbf{A}_{\mathbf{x}\mu}$ being  the vector gauge field.

As the dominant contribution to the Raman response from the spin ice Hamiltonian comes from the spinon excitations \cite{Fu2017}, in the following  we focus on the  spinon part of the Hamiltonian only. In this case,    the expressions (\ref{eqn:spinon-p}) and (\ref{eqn:spinon-m}), can be  further simplified into $\tilde{\tau}_{\mathbf{x},\mu}^+=\psi_\mathbf{x}^\dagger \psi_{\mathbf{x}+\hat{\boldsymbol{\mu}}}/2$ and $\tilde{\tau}_{\mathbf{x},\mu}^-=\psi_{\mathbf{x}+\hat{\boldsymbol{\mu}}}^\dagger\psi_\mathbf{x}/2$.  
Then using the bosonic representation of the quantum XY rotor model  discussed in  \cite{Hao2014}
    \begin{align}
        Q_\mathbf{x}=\,&d^\dagger_{\mathbf{x}} d_{\mathbf{x}}-b_{\mathbf{x}}^\dagger b_{\mathbf{x}},\\
        Q^2_{\mathbf{x}}\approx\, &d^\dagger_{\mathbf{x}} d_{\mathbf{x}}+b^\dagger_{\mathbf{x}} b_{\mathbf{x}},\\
        \psi_{\mathbf{x}}\approx\,&d_{\mathbf{x}}+b^\dagger_{\mathbf{x}},
    \end{align}
    we obtain the quadratic bosonic Hamiltonian for spinons:
\begin{widetext}
    \begin{align}\label{SpinonHamiltonianwide}
        \mathcal{H}_{\mathrm{Nd}
        %,(A,B)
        }^s=&\frac{1}{2}\left(\tilde{J}_{z} -h_\mathrm{loc}\right)\sum_{\mathbf{k}}d^\dagger_{\mathbf{k}}d_{\mathbf{k}}+\frac{1}{2}\left(\tilde{J}_z +h_{\mathrm{loc}} \right)\sum_{\mathbf{k}}b^\dagger_{\mathbf{k}}b_{\mathbf{k}}\nonumber\\&-\frac{\tilde{J}_{xy}}{16}\sum_{\mathbf{k},\alpha\neq\beta}^{x,y,z}\cos\frac{k_\alpha}{2}\cos\frac{k_\beta}{2}\left(d^\dagger_{\mathbf{k}}d_\mathbf{k}+b_{-\mathbf{k}}b^\dagger_{-\mathbf{k}}+b^\dagger_{\mathbf{k}}d^\dagger_{-\mathbf{k}}+d^\dagger_{\mathbf{k}}b^\dagger_{-\mathbf{k}}+\mathrm{h.c.}\right).
     \end{align}
\end{widetext}

This Hamiltonian can be  diagonalized with the Bogoliubov transformation, which gives the spectrum for the spinon excitations
\begin{align}
    \omega(\mathbf{k})=
    \frac{1}{2}\left(h_{\mathrm{loc}}\pm \tilde{J}_{z}\sqrt{1-\frac{\tilde{J}_{xy}}{2\tilde{J}_{z}}\sum_{\alpha\neq\beta}^{x,y,z}\cos\frac{k_\alpha}{2}\cos\frac{k_\beta}{2}}\right),
\end{align}
 where $\alpha,\beta={x,y,z}$ are the three  cubic directions.

\subsection{ Raman responses of two-spinon excitations}

Next, we compute the intensity of the magnetic Raman peak M1, which is experimentally observed around 14 meV, by utilizing the Raman operator for quantum spin ice, as derived in \cite{Fu2017}.  Using this framework, we  compute the contributions of spinon excitations into Raman response under the assumption that the effective Hamiltonian (\ref{SpinonHamiltonianwide}) captures  the dynamics of Nd magnetic moments. This approach enables us to directly compare the theoretical spectrum with experimental data, offering insights into the consistency of the two-spinon continuum interpretation with the observed M1 peak.

The Raman operator for the quantum spin ice can be written as  \cite{Fu2017}
    \begin{align}
        &\mathcal{R}_{\mathrm{Nd}}=
        \sum_{\langle ij\rangle}\\\nonumber&\left[(\mathbf{e}_{\mathrm{in}}\cdot\boldsymbol{\mu}_i)(\mathbf{e}_{\mathrm{out}}\cdot\boldsymbol{\mu}_i)+(\mathbf{e}_{\mathrm{in}}\cdot\boldsymbol{\mu}_j)(\mathbf{e}_{\mathrm{out}}\cdot\boldsymbol{\mu}_j)\right]\mathcal{H}_{\mathrm{Nd-Nd},ij},\nonumber
    \end{align}    
where ${\boldsymbol\mu}_i$ and ${\boldsymbol \mu}_j$ 
are defined in Fig.3 d in the main text, and 
$\mathcal{H}_{\mathrm{Nd-Nd},ij}$ are renormalized 
interaction between Nd moments  in (\ref{eqn:renormalized}). To match the energy and width
of the M1 peak, we set $\tilde{J}_z=12.0$ meV and $\tilde{J}_{xy}=-3.0$ meV
and assume that just below 33 K
the local field is $h_{\mathrm{loc}}=12.0$ meV.
We also take the same incoming and outgoing light polarization direction as we compute the Raman response for the one-magnon excitation in the [111] plane. The  resulting intensity is shown in Fig.\ref {fig:M1-peak-theory}.

\begin{figure}[t!]
    \centering
    \includegraphics[width=0.9\linewidth]{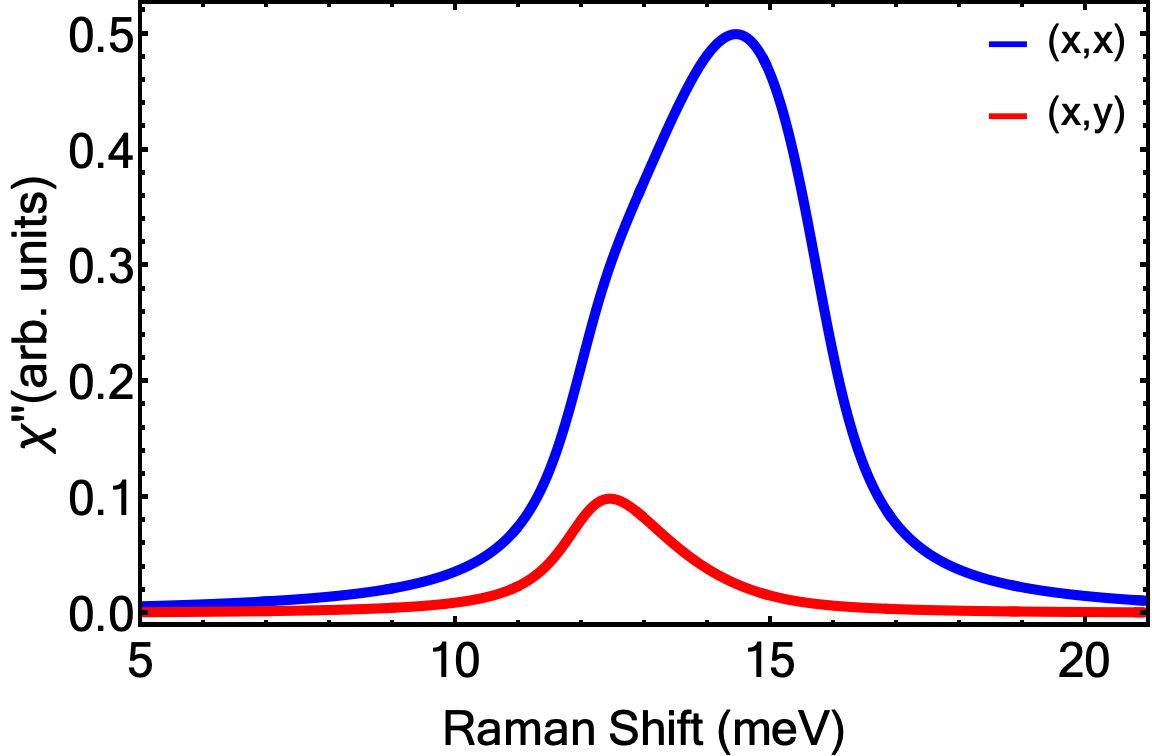}
    \caption{Computed Raman responses from two-spinon excitation continuum in both parallel and crossed channel.}
    \label{fig:M1-peak-theory}
\end{figure}

Additionally, we have successfully simulated the temperature-dependent behavior of the M1 peak
  presented in  Fig.1 of the main text 
  by changing the local field strength $h_{\mathrm{loc}}$. 
 Namely, as  the magnetic moment on the Ir ion $\langle m_{\rm Ir}\rangle$ increases with lowering of temperature,
 the local field acting on the Nd ion also becomes larger. Thus, by increasing $h_{\mathrm{loc}}$ from $12$ meV to $16$ meV, we mimic the temperature evolution of the two-spinon Raman response with lowering of temperature.  Computed intensity  plot is presented in Fig.~\ref{fig:h_loc}.

}

\begin{figure}[h!]
    \centering
    \includegraphics[width=0.9\linewidth]{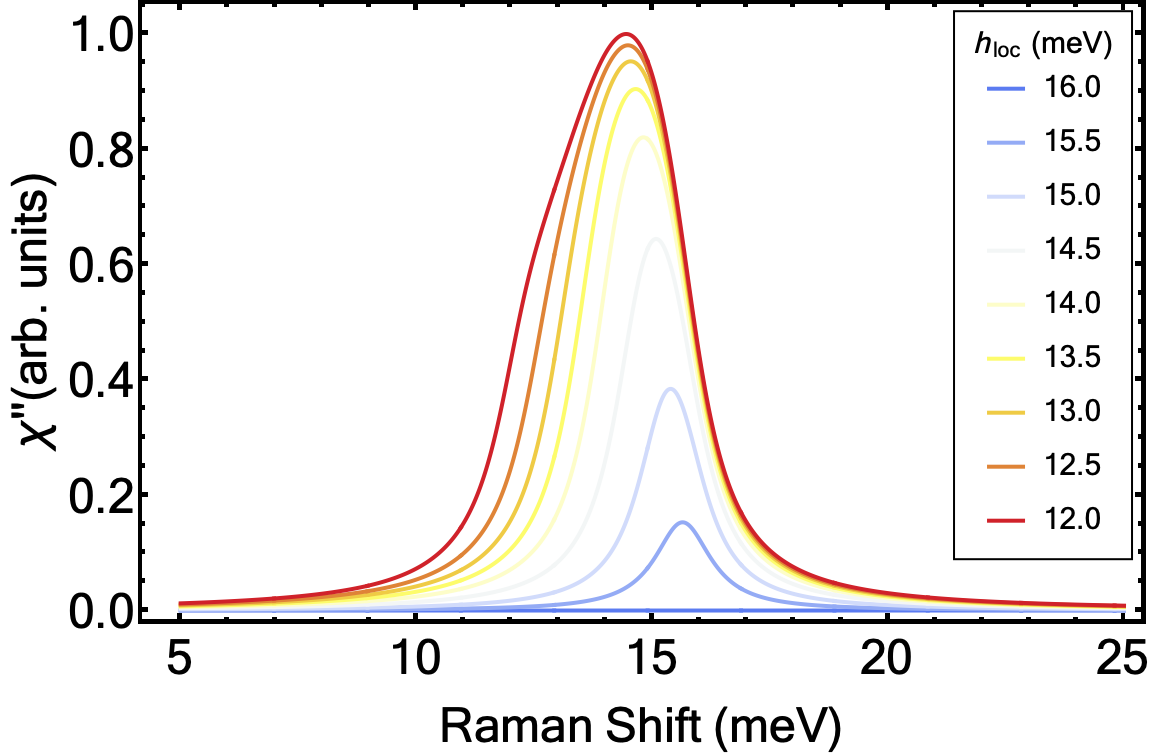}
    \caption{Temperature dependence of M1 peak mimicked by increasing the local field $h_{\mathrm{loc}}$ (larger local field corresponds to lower temperature).} 
    \label{fig:h_loc}
\end{figure}

\section{Fitting the spectra}

\begin{figure*}[!htb]
    \centering
    \includegraphics[width=\textwidth]{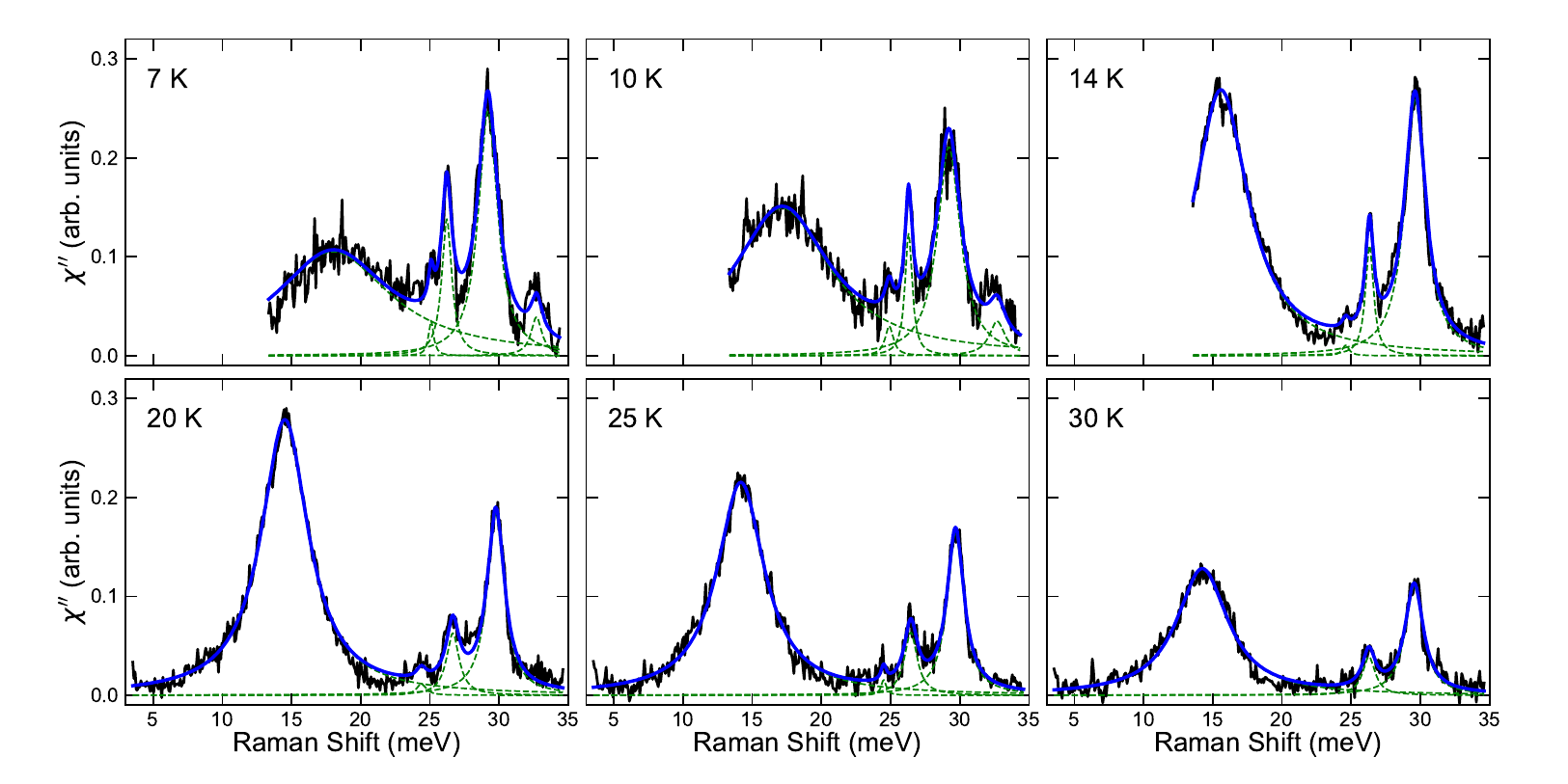}
    \caption{Fitting results of the Raman spectra in $(x,x)$ polarization for the magnetic excitations M1, M2, M3, and M4. Solid black lines denote the experimental spectra with subtracted electronic contribution $\chi''(\omega) - \chi''_0(\omega)$. Dashed green lines denote the Lorentzian fits for each peak. Solid blue curves denote the fitted curves.}
    \label{fig:fit_magnetic}
\end{figure*}

\begin{figure*}[!htb]
    \centering
    \includegraphics[width=\textwidth]{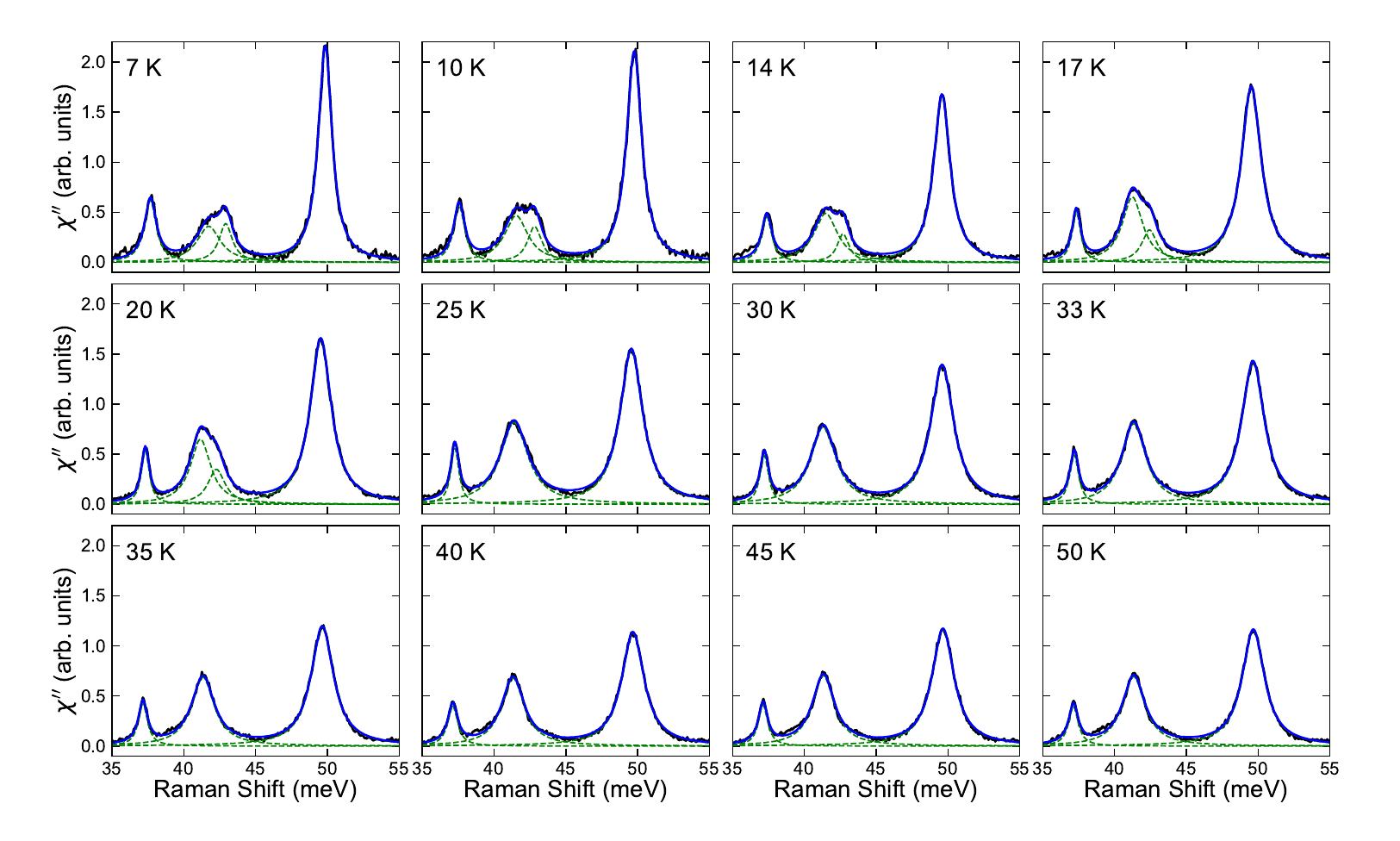}
    \caption{Fitting results of the Raman spectra in $(x,x)$ polarization for the phonon excitations $T_{2g}^{(1)}$, $E_g$, and $T_{2g}^{(2)}$. Solid black lines denote the experimental spectra with subtracted electronic contribution $\chi''(\omega) - \chi''_0(\omega)$. Dashed green lines denote the Lorentzian fits for each peak. Solid blue curves denote the fitted curves.}
    \label{fig:fit_phonons}
\end{figure*}

To obtain the temperature dependence of the parameters of the observed excitations presented in the figures in the main text of the manuscript we fitted the spectra with a function discussed in the Methods section. In the Fig.~\ref{fig:fit_magnetic} and \ref{fig:fit_phonons} we present a comparison of the fitting curves with the experimental data.

\end{document}